\documentclass[final,11pt,times,3p,freqn]{elsarticle}

\usepackage{geometry}
\usepackage[utf8]{inputenc} 
\usepackage[T1]{fontenc}    
\usepackage{hyperref}       
\usepackage{url}            
\usepackage{booktabs}       
\usepackage{amsfonts}       
\usepackage{nicefrac}       

\usepackage{amssymb}
\usepackage{amsmath}
\usepackage{amsthm}

\hypersetup{%
  colorlinks=true,
  pdfborder={0 0 0 },
}

\graphicspath{{./figs}}

\usepackage{nameref}
\usepackage{natbib}
\usepackage[separate-uncertainty=true,multi-part-units=single]{siunitx}
\usepackage{xcolor}
\usepackage{amsmath}
\usepackage[capitalize]{cleveref}
\crefformat{appendix}{#2#1#3}

\usepackage{tabularx}  
\newcolumntype{C}{>{\centering\arraybackslash}X}
\newcolumntype{R}{>{\raggedleft\arraybackslash}X}
\newcolumntype{L}{>{\raggedright\arraybackslash}X}

\usepackage[small]{caption}
\usepackage{xspace}
\usepackage{multirow}

\usepackage{lineno}
\usepackage{balance}

\usepackage[table]{xcolor}
\definecolor{tbH}{RGB}{0,33,84}
\definecolor{tb1}{RGB}{216,226,244}
\usepackage{booktabs}
\newcommand{\head}[1]{\textcolor{white}{\textbf{#1}}}

\newcommand{\pdif}[2]{\frac{\partial #1}{\partial #2}}
\newcommand{\fdif}[2]{\frac{\delta #1}{\delta #2}}
\newcommand{\plap}[2]{\frac{\partial^2 #1}{\partial {#2}^2}}

\newcommand{\bol}[1]{\boldsymbol{#1}}
\newcommand{\paren}[1]{\left( #1 \right)}
\newcommand{\Rey}{\mathit{Re}}

\begin{document}
\begin{frontmatter}

\title{GIMLET: \underline{G}eneralizable and \underline{I}nterpretable \underline{M}odel \underline{L}earning through \underline{E}mbedded \underline{T}hermodynamics}

\author[1,2]{Suguru Shiratori\corref{cor}}
\ead{sshrator@tcu.ac.jp}
\author[2]{Elham Kiyani}
\author[2]{Khemraj Shukla}
\author[2]{George Em Karniadakis}

\cortext[cor]{Corresponding author}
\address[1]{Department of Mechanical Systems Engineering, Tokyo City University, 1-28-1 Tamazutsumi, Setagaya-ku, 158-8557 Tokyo, Japan}
\address[2]{Division of Applied Mathematics, Brown University, 182 George Street, Providence, 02912, RI, USA}

\begin{abstract}
We develop a data-driven framework for discovering constitutive relations in models of fluid flow and scalar transport. Under the assumption that velocity and/or scalar fields are measured, our approach infers unknown closure terms in the governing equations as neural networks (gray-box discovery). The target to be discovered is the constitutive relations only, while the temporal derivative, convective transport terms, and pressure-gradient term in the governing equations are prescribed.
The formulation is rooted in a variational principle from non-equilibrium thermodynamics, where the dynamics is defined by a free-energy functional and a dissipation functional. The unknown constitutive terms arise as functional derivatives of these functionals with respect to the state variables.
To enable a flexible and structured model discovery, the free-energy and dissipation functionals are parameterized using neural networks, while their functional derivatives are obtained via automatic differentiation. This construction enforces thermodynamic consistency by design, guaranteeing non-negative entropy production and, in the absence of non-variational (convective) contributions, monotonic decay of the total free energy. The resulting method, termed \emph{GIMLET} (Generalizable and Interpretable Model Learning through Embedded Thermodynamics), avoids reliance on a predefined library of candidate functions, unlike sparse regression or symbolic identification approaches.
The learned models are generalizable in that functionals identified from one dataset can be transferred to distinct datasets governed by the same underlying equations. Moreover, the inferred free-energy and dissipation functions provide direct physical interpretability of the learned dynamics. The framework is demonstrated on several benchmark systems, including the viscous Burgers equation, the Kuramoto--Sivashinsky equation, and the incompressible Navier--Stokes equations for both Newtonian and non-Newtonian fluids.
\end{abstract}

\begin{keyword}
Structure-preserving \sep
Physics-informed neural networks \sep
Gray-box discovery \sep
Functional derivatives \sep
Non-equilibrium thermodynamics
\end{keyword}

\end{frontmatter}

\section{Introduction}\label{sec-intro} 
Discovering the governing laws of physical systems is one of the most noble and impactful tasks in science and engineering. For the design and development of various products and infrastructures, numerical simulations based on partial differential equations (PDEs) have been essential tools; however, they often exhibit considerable discrepancies from real observational data. One possible cause of this discrepancy is misspecification of parameters, such as physical properties or initial conditions, in which case simulations can be improved through parameter tuning or data assimilation.
More fundamentally, parts of the governing equations (GEs) used in simulations may be missing or misspecified. In such cases, one must revisit the underlying physics and reformulate the GEs, a process that typically requires enormous effort. For instance, in fluid dynamics, determining constitutive laws for non-Newtonian fluids or viscoelastic materials remains a longstanding challenge.

Classical approaches for physical model discovery have relied on expert-derived equations grounded in physical principles, phenomenology, and empirical observations~\cite{truesdell2004non,gurtin2010mechanics,marsden1994mathematical,Onsager1931a}. 
The rapidly growing availability of high-fidelity experimental data, together with advances in machine learning, has enabled data-driven discovery of physical models and established it as an emerging paradigm~\cite{lee2020coarse}.
Previously proposed methodologies span a spectrum from black-box to gray-box models. In this work, we consider gray-box discovery and focus on two representative families, neural operator style models and symbolic regression, and compare them through the lenses of generalizability and interpretability.

Neural operator style models aim to learn mappings between function spaces, such as the evolution of fields over time.
Representative methods include Neural ODE~\cite{Chen2018}, Hamiltonian Neural Networks~\cite{Greydanus2019}, Fourier Neural Operator~\cite{Li2020}, DeepONet~\cite{Lu2021}, and many others~\cite{Lusch2018,Champion2019,Pfister2019,Yuan2019}.
These approaches are largely black-box and often offer limited interpretability. In terms of generalizability, they typically require training on diverse data spanning many operating conditions, which can be time-consuming.

In contrast, symbolic regression (SR) methods lie closer to the white box end of the spectrum and aim to recover explicit governing equations from data. SR comprises methods that discover mathematical expressions that best fit a given dataset~\cite{schmidt2009symbolic,stoutemyer2013can}. Early successes in data-driven equation discovery were achieved with genetic programming~\cite{Bongard2007,Schmidt2009}, still the most widely used family of SR methods. There, candidate models are represented as expression trees whose internal nodes are elementary operators and whose leaves are variables or constants, so that every candidate carries an explicit analytical form. A population of such trees is improved iteratively through selection, mutation, and crossover, scored by agreement with the data. In recent work by~\cite{cranmer2023interpretable}, both the performance and computational efficiency of symbolic regression were further enhanced through an island-based evolutionary framework. In this approach, populations were partitioned into multiple islands, each characterized by distinct operator sets and mutation strategies. Genetic algorithm operations, including mutation and crossover, were performed both within individual islands and across islands, promoting population diversity, improved exploration of the search space, and more effective convergence toward interpretable symbolic models.

Building on this insight, sparse modeling techniques provided a more tractable and scalable route to interpretable discovery. 
The Sparse Identification of Nonlinear Dynamics (SINDy) framework of~\citet{Brunton2016} advanced the field by selecting parsimonious terms from large candidate libraries. This idea was later extended to PDEs through sparse dynamics~\cite{Schaeffer2013}, sparse optimization~\cite{Schaeffer2017}, and data-driven PDE discovery in real systems~\cite{Rudy2017}.
Subsequent work addressed noise, model complexity, and high-dimensional settings, including PDE discovery in heterogeneous fields~\cite{Berg2019} and hybrid neural sparse architectures such as DeepMoD~\cite{Both2021}.
All the SR approaches yield explicit and interpretable models, but their ability to generalize is not always guaranteed.
Both genetic programming and sparse identification typically rely on a predefined library of candidate functions or terms.
If the appropriate function is not represented in this library, the method may select ad-hoc functions that fit the training data but do not extrapolate, limiting generalization beyond the observed regime.

Physics-Informed Neural Networks (PINNs)~\cite{Raissi2019} are a groundbreaking method and a main pillar of the scientific machine learning (SciML) field, which involves data-driven physics model discovery.
PINNs can seamlessly integrate physical laws as an additional loss function (physics loss) to the conventional supervised loss function in neural networks, by evaluating errors in PDEs by Automatic Differentiation (AD).
PINNs have been applied to numerous problems, which include thermo-fluidics models of crystal growth~\cite{Takehara2023}, two-phase flow problems~\cite {Qiu2022}, or liquid film flows~\cite{Nakamura2022}.
Also, many extensions of PINNs have been proposed, such as gradient-enhanced PINNs~\cite{Yu2022}, Generative Adversarial PINNs (GA-PINNs)~\cite{Li2022}, Bayesian PINNs (B-PINNs)~\cite{Yang2021,Kiyani2024}, PINNs coupled with symbolic regression~\cite{Kiyani2023,Zhang2024}, and coupling PINNs with numerical solvers for multiphysics problems~\cite{Shukla2025}.

The concept of adding physics loss and the technique of AD are widely applied in the field of physics model discovery.
\citet{Zou2024} have proposed a method to correct the misspecified physical model by combining PINNs and an additional neural network.
ViscoelasticNet, proposed by \citet{Thakur2024}, enabled data-driven parameter identification in constitutive models for viscoelastic fluids.
For the sparse identifications, PINN-SR~\cite{Chen2021} successfully identified GEs from scarce and noisy data by evaluating spatial derivatives by ADs, which had been discretized in PDE-FIND~\cite{Rudy2017}.
By introducing the physics losses, the aforementioned Neural Operators can be `physics-informed' and they become gray-box discovery, such as Physics-informed DeepONets \cite{Wang2021} or Physics-informed Neural Operators~\cite{Li2021,Zhong2025}.
This way, the generalizability and interpretability of Neural Operators can be improved; nevertheless, they still require large datasets for training.

As reviewed so far, existing methods for physical model discovery typically fall short in at least one of three aspects: generalizability, interpretability, or independence from a predefined library.
We propose a new methodology that aims to achieve all three objectives. To this end, we leverage thermodynamics and a variational principle.
%
Thermodynamics is among the most universal and transferable descriptions of physical systems.
Enforcing thermodynamic consistency can lead to more robust constitutive laws, stable numerical schemes, and physically meaningful discovered models~\cite{Cueto2023}.
Motivated by this idea, several machine learning frameworks have been proposed, including Thermodynamics-based Artificial Neural Networks (TANNs)~\cite{Masi2021} and thermodynamically consistent physics-informed neural networks~\cite{Patel2022}.
Motivated by the same goal of enforcing thermodynamic consistency in a general and systematic way, the General Equation for Non-Equilibrium Reversible Irreversible Coupling (GENERIC) framework was developed as a universal description of macroscopic dynamics. In GENERIC, the evolution equations are constructed from energy and entropy potentials, providing a unified structure that couples reversible and irreversible processes and can be viewed as a thermodynamically grounded extension of the Ginzburg--Landau equation~\cite{Grmela1997,Oettinger1997}.
Based on the GENERIC concept, \citet{Zhang2022} proposed the GENERIC formalism-informed neural networks (GFINNs) for physical model discovery. However, the applicability of GFINN has so far been investigated primarily for ordinary differential equations, leaving extensions to PDE settings open.

According to the variational principle, also called the principle of least action, realizable motions correspond to the stationary points of an action functional.
Onsager's variational principle~\cite{Onsager1931a,Onsager1931}, which formulated the time evolution of the free energy and dissipation potentials, can lead to thermodynamically consistent governing equations for non-equilibrium dynamics~\cite{Van2020}.
Building on this principle, \citet{Huang2022} proposed the Variational Onsager Neural Networks (VONNs), which aim to discover the free energy and dissipation potentials as neural networks.
Although VONNs are a sophisticated and novel methodology, their application is limited to systems whose dynamics can be fully described by free energy and dissipation potentials.
This restriction makes direct application to fluid flow challenging.
For instance, applying Onsager's variational principle to the Newtonian fluid in a straightforward way yields only a linearized form of the Navier--Stokes equation~\cite{Takaki2007}.
In addition, the VONNs framework assumes that both the state variables and process variables are available as training data, which is often unrealistic for practical experimental observations.

In summary, our study aims to resolve the aforementioned issues in the previously proposed methods and to formulate a new methodology that can be applied to fluid flow problems.
In the remainder of this paper, the proposed methodology is formulated in \cref{sec-method}, followed by the validation of the method in \cref{sec-validation}.
The discussion and conclusions are provided in \cref{sec-conclusion}.

\section{Methodology}\label{sec-method}
This section presents our thermodynamically consistent, 
library-free framework for discovering missing terms in continuum PDEs.

We consider physical law discovery in continuum mechanics, where spatiotemporal field variables are evolving according to partial differential equations (PDEs).
We assume that part of the governing equations is known and focus only on the missing terms (gray-box discovery).
To represent these unknown contributions, an additional free energy function $g$ and dissipation function $\theta$ are introduced.
Basic forms of governing equations (GEs) are derived through the variational principle.
In the resulting GEs, the unknown terms are expressed as functional derivatives of $g$ and $\theta$.
To learn $g$ and $\theta$ from data, we employ physics-informed neural networks (PINNs).
Using PINNs, the functional derivatives for unknown functions can be calculated through automatic differentiation.
The total loss function is defined as the sum of the data loss and physics loss.
The data loss represents an error between observed data and the PINN predictions, whereas the physics loss enforces the derived governing equations by penalizing their residuals.
Minimizing this loss simultaneously fits the observed dynamics and identifies the missing terms through the learned functions $g$ and $\theta$.

\subsection{Problem formulation}
The dynamics of a viscous two-component fluid are considered.
Let $\rho$ and $\phi$ be the total mass density and the mass fraction of one component, respectively.
The mass conservation can be given as
\begin{gather}
  \partial_t \rho + \nabla\cdot\left( \rho \bol{u}\right) = 0, \\
  \rho\left( \partial_t \phi + \left(\bol{u}\cdot\nabla\right)\phi \right) = - \nabla\cdot\bol{j}_a, \label{eq-phiTrans}
\end{gather}
where $\bol{u}$ and $\bol{j}_a$ are the velocity fields and the diffusive flux, respectively.
The chemical potential is defined as the Gibbs energy per unit mass, and $\mu\paren{\phi}$ is the difference of chemical potential between two components.
In addition to the free energy due to the fluid motion, the free energy due to the two-component mixture is defined as $g(\phi)$, and the form of $g\paren{\phi}$ is inferred from the observed data.
The total additional free energy of this system, $G$, and the chemical potential $\mu$ are defined as
\begin{equation}
   G[\phi] \equiv \int_V g(\phi) dV, \qquad
   \mu \equiv \frac{1}{\rho}\frac{\delta G}{\delta \phi},
\end{equation}
where $\delta / \delta \phi$ is the functional derivative.
The specific internal energy $\epsilon$ is a function of the density $\rho$, the entropy $s$, and the mass fraction $\phi$ as
\begin{equation}
   p   \equiv \rho^2 \left( \pdif{\epsilon}{\rho} \right)_{s,\phi}, \qquad
   T   \equiv \left( \pdif{\epsilon}{s}   \right)_{\rho,\phi},      \qquad
   \mu \equiv \left( \pdif{\epsilon}{\phi} \right)_{s,\rho},
\end{equation}
where the subscripts indicate variables fixed in the respective partial differentiations.
Here $\epsilon$ denotes the internal energy of the base fluid, which does not include the additional mixing free energy $g\paren{\phi}$; the latter enters the formulation only through $G[\phi]$.
From thermodynamics, the total differentiation of $\epsilon$ can be written as
\begin{equation}
   d\epsilon =  \frac{p}{\rho^2}d\rho + Tds + \mu d\phi.   \label{eq-freeEnergy}
\end{equation}

For dissipative processes, the time evolution of entropy can be written as
\begin{equation}
   \rho T \left( \pdif{s}{t} + \bol{u}\cdot\nabla s \right)  - \theta
   + \nabla\cdot \left( \bol{q} + \mu\bol{j}_a \right)= 0, \label{eq-constraintNH}
\end{equation}
where $\bol{q}$ is the heat flux, and $\theta$ is the dissipation function, which expresses how the frictional force converts the energy into heat.
The total dissipation potential is defined as
\begin{equation}
   \Theta[\bol{u},\phi] \equiv \int_V \theta(\nabla\bol{u},\nabla\phi) dV.
\end{equation}

In the proposed method, the contribution of additional free energy, which corresponds to $\mu$ in \cref{eq-freeEnergy}, is assumed to be time-independent.
This means that the present form of this method is limited to discovering the time-independent constitutive laws. This framework can be extended for the constitutive laws that involve memory effect, for instance, viscoelastic fluids or turbulent closures. To this end, the free energy contribution \cref{eq-freeEnergy} must be modified accordingly so that the memory effect can be expressed properly.

\subsection{Variational principle and basic form of the governing equations}
Next, we derive the governing equations by considering the principle of least action.
The variational principle formulated by \citet{Fukagawa2010,Fukagawa2012} is adopted, which can derive the Navier--Stokes equation including the inertia term.
In the following, an overview of the derivation is described.
The details of the derivation are provided in \ref{app-VP}.
We define the Lagrangian density as
\begin{equation}
   \mathcal{L}(\rho, \bol{u}, s, \phi)  \equiv \rho \left\{ \frac{1}{2}\bol{u}^2  - \epsilon(\rho,s,\phi) \right\},
\end{equation}
and the action path as
\begin{equation}
I = \int_{t_\text{ini}}^{t_\text{fin}} dt \int_V dV
\{ \mathcal{L} - \lambda U\paren{\bol{u},\rho,s}   \},   \label{eq-action}
\end{equation}
where $U\paren{\bol{u}, \rho, s}$ is the constraint and $\lambda$ is the Lagrange multiplier.
By solving the stationary condition of \cref{eq-action}, the momentum equation is obtained.
From the second law of thermodynamics, $\bol{j}_a \cdot \nabla\mu \leq 0$ must be satisfied.
Assuming the linear phenomenological law, we can express $\bol{j}_a$ as
\begin{equation}
 \bol{j}_a = -\frac{D}{T} \nabla \mu,
\end{equation}
where $D (>0)$ is a positive constant called a diffusion coefficient.
With the expression of $\mu$ with the free energy functional, the transport equation for $\phi$ can be written as
\begin{equation}
\partial_t \phi + \bol{u}\cdot\nabla\phi
= \nabla\cdot \left( \frac{D}{T} \nabla \frac{\delta G}{\delta \phi} \right).
\end{equation}
In this study, we consider the isothermal condition; thus, we replace $D/T$ with a constant $M$.
From what has been formulated so far, the basic form of GEs can be written as
\begin{subequations}
\label{eq-GE1}
\begin{gather}
\rho \paren{ \pdif{\bol{u}}{t} + \bol{u}\cdot\nabla \bol{u} }
= -\nabla p +  \paren{\nabla\phi}\frac{\delta G}{\delta \phi}  - \frac{\delta \Theta}{\delta \bol{u}}, \\
\pdif{\phi}{t} + \bol{u}\cdot\nabla\phi
= \nabla\cdot \paren{ M \nabla \frac{\delta G}{\delta \phi} },
\end{gather}
\end{subequations}
and our task is to discover the free energy function $g$ and the dissipation function $\theta$ from the observation data.

\subsection{Architecture of GIMLET}
\begin{figure}
    \centering
    \includegraphics[width=\textwidth]{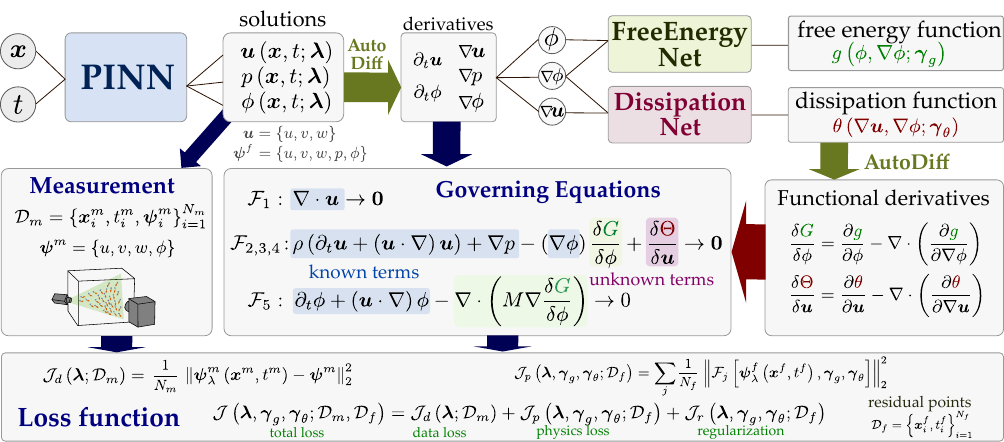}
\caption{\label{fig-structure}
Schematic architecture of \emph{GIMLET} for data-driven discovery of constitutive models in flow and transport PDEs.
The PINN predicts the solution fields $\psi^f_{\bol{\lambda}}$ from spatiotemporal inputs $\{ \bol{x}, t \}$, while FreeEnergyNet and DissipationNet parameterize the free-energy and dissipation functions.
Automatic differentiation is used to evaluate functional derivatives, and training is performed by minimizing combined data and physics losses.
The pressure field is assumed not to be measured, and the data loss is evaluated without pressure data.
For the problem of fluid flow only, the scalar field $\phi$ and FreeEnergyNet are not used.}
\end{figure}

\Cref{fig-structure} shows the architecture of the proposed method.
The total framework consists of three neural networks: a PINN, FreeEnergyNet, and DissipationNet.
The PINN, which is trainable through parameters $\bol{\lambda}$, learns the mapping of input coordinates $\{\bol{x}, t\}$ to the solution $\bol{\psi}^f_{\bol{\lambda}} = \{u,v,w,p,\phi\}$.
The FreeEnergyNet and DissipationNet learn the mapping of solutions to the free energy function $g$ and the dissipation function $\theta$, respectively. 
These neural networks are trainable through $\bol{\gamma}_g$ and $\bol{\gamma}_\theta$, respectively.
The total loss function $\mathcal{J}$ is composed of the data loss $\mathcal{J}_d$, the physics loss $\mathcal{J}_p$, and the regularization terms $\mathcal{J}_r$.
The data loss and physics loss are calculated using the measurement dataset $\mathcal{D}_m$ and residual points $\mathcal{D}_f$, respectively.
By optimizing three neural networks simultaneously, the PINN is trained to infer the full solution field and the missing terms. 
The pressure field is assumed not to be measured, and the data loss is evaluated without pressure data.
For the problem of fluid flow only, the scalar field $\phi$ and FreeEnergyNet are not used. 

\subsubsection{Physics-informed neural networks (PINNs)}
The PINNs methodology enables imposing the physical laws in the machine learning framework as soft constraints through a loss function.
We use the PINN for the mapping of the solution field $\bol{\psi}^f_\lambda$ from spatiotemporal coordinates $\{\bol{x}, t\}$, and for taking spatial and time derivatives through AD. 
Among a lot of variations of PINNs, the original plain PINN \cite{Raissi2019} is employed in this study.

\subsubsection{FreeEnergyNet}
The free energy function $g\paren{\phi}$ is expressed by the FreeEnergyNet.
Since only derivatives of $g\paren{\phi}$ are required in the GEs, the FreeEnergyNet is composed of an Integrable Neural Network (INN) \cite{Teichert2019}.
The INN is a fully-connected network whose gradient can be evaluated exactly by a single forward pass, so that both $f$ and $\partial f/\partial \xi_k$ are available without backpropagation.
We consider the standard fully-connected neural network associated with a multi-variable function $f\paren{\bol{\xi}}= f\paren{\xi_1, \xi_2, \ldots, \xi_m}$:
\begin{equation}
    \bol{y}^\ell = \bol{W}^\ell \bol{z}^{\ell-1} + \bol{b}^\ell, \qquad
    \bol{z}^\ell = \sigma \paren{\bol{y}^\ell}, \qquad
    f = \bol{W}^{n+1} \bol{z}^n, \qquad
    \bol{z}^0 = \tilde{\bol{\xi}} \label{eq-INN1}
\end{equation}
where $\tilde{\bol{\xi}}$ is normalized input, $\bol{z}^\ell$ is the activated values of $\ell$-th hidden layer, $\bol{y}^\ell$ is the value before activation, $\sigma$ is activation function, $\bol{W}^\ell$ and $\bol{b}^\ell$ are the weights and biases.
\Cref{eq-INN1} can be analytically differentiated as
\begin{equation}
    \bol{\zeta}_{k}^\ell = \sigma'\paren{\bol{y}^\ell} \bol{W}^\ell \bol{\zeta}_{k}^{\ell-1}, \qquad
    \pdif{f}{\bol{\xi}_k} = \bol{W}^{n+1} \bol{\zeta}_{k}^n, \label{eq-INN2}
\end{equation}
where $\sigma'$ is the derivative of activation function, $\bol{\zeta}_{k}^\ell$ is additional activation units.
During the training, the derivative $\partial f / \partial \xi_k$ is used in the loss function, whereas the original function $f$ is used for the interpretation of the discovered model.
Although the derivative of the function can also be calculated using Automatic Differentiation (AD), a set of forward and backward propagation is required.
By use of the INN, the derivative can be obtained by a single forward propagation.
To strictly enforce $f\paren{\bol{0}} = 0$, the following offset function is used:
\begin{equation}
    f\paren{\bol{\xi}} = \tilde{f}\paren{\bol{\xi}} - \tilde{f}\paren{\bol{0}},
\end{equation}
where $\tilde{f}$ is the output of the INN.
For the activation functions $\sigma$ and its derivative $\sigma'$, the {\em softplus} and {\em sigmoid} functions are employed in this study.

\subsubsection{DissipationNet}
The dissipation function $\theta\paren{\bol{u}}$ is expressed by the DissipationNet, which is composed of a Convex Integrable Neural Network (CINN) \cite{Amos2017}.
The CINN is a modified INN, on which the convexity condition is strictly enforced.
According to \cite{Huang2022}, the following CINN is introduced.
\begin{equation}
    \bol{z}^{\ell+1} = \sigma \paren{ \bol{W}^\ell_y \bol{z}^\ell + \bol{W}_w^\ell \tilde{\bol{\xi}} + \bol{b}^\ell }, \qquad
    \bol{z}^0 = \tilde{\bol{\xi}}, \label{eq-CINN1}
\end{equation}
where $\bol{W}^\ell_y$ and $\bol{W}^\ell_w$ are the weight matrices for hidden layers and passthrough layers, respectively.
At the connection between the input and the first hidden layer, only the passthrough connection is considered, letting $\bol{W}_y^0 = \bol{0}$.
To maintain the convexity with respect to $\bol{\xi}$, the weights $\bol{W}^\ell_y$ must be non-negative and the activation function $\sigma$ must be both convex and non-decreasing.
As in the INN, the offset $\theta\paren{\bol{\xi}} = \tilde{\theta}\paren{\bol{\xi}} - \tilde{\theta}\paren{\bol{0}}$ is applied so that $\theta\paren{\bol{0}} = 0$.
Together with the convexity and $\nabla\theta\paren{\bol{0}} = \bol{0}$, this guarantees $\theta \geq 0$, and hence non-negative entropy production, irrespective of the training outcome.

\subsubsection{Functional derivatives}
$\delta G/\delta \phi$ and $\delta \Theta / \delta \bol{u}$ appearing in \cref{eq-GE1} are the functional derivatives, which can be calculated through AD.
In general, the free energy and dissipation functions might depend on the derivatives of the solution field $\bol{u}$ and/or $\phi$ as
\begin{subequations}   
\begin{gather}
G\left[ \phi\paren{\bol{x}} \right] = \int_\Omega g \paren{ \bol{x}, \phi\left(\bol{x}\right), \nabla \phi\left(\bol{x}\right), \ldots, \nabla^{(n)} \phi\left(\bol{x}\right) }  d\Omega, \\
\Theta\left[ \bol{u}\paren{\bol{x}} \right] = \int_\Omega \theta \paren{ \bol{x}, \bol{u}\left(\bol{x}\right), \nabla \bol{u}\left(\bol{x}\right), \ldots, \nabla^{(n)} \bol{u}\left(\bol{x}\right) }  d\Omega.
\end{gather}
\end{subequations}
For these forms, the functional derivatives can be written by using series expansions as
\begin{subequations}
\label{eq-FD2}
\begin{align}
\fdif{G}{\phi} &= \pdif{g}{\phi} - \nabla\cdot\pdif{g}{\left( \nabla \phi \right)} + \nabla^2\cdot\pdif{g}{\left( \nabla^2 \phi \right)} + \cdots + \left( -1 \right)^{(n)} \nabla^{(n)} \cdot \pdif{g}{\left( \nabla^{(n)} \phi \right)}, \\
            &= \pdif{g}{\phi} + \sum_{i=1}^{n} \left( -1 \right)^{(i)} \nabla^{(i)}  \cdot \pdif{g}{\left( \nabla^{(i)} \phi \right)}.
\end{align}
\end{subequations}
For most cases of physical phenomena, truncation up to the second term is sufficient.
All the derivatives in \cref{eq-FD2} can be numerically evaluated by AD.
The first term $\partial g/\partial \phi$ can be obtained without AD by using the INN or the CINN, which outputs the derivatives directly.
For the second term $\nabla\cdot \paren{\partial g / \partial \paren{\nabla\phi} }$, the gradient $\nabla\phi$ can be calculated by AD through the PINN.
After $\partial g / \partial \paren{\nabla\phi}$ is evaluated by the FreeEnergyNet, its divergence can be calculated by AD through both the PINN and FreeEnergyNet.
One of the novelties in this study is that the evaluation of these functional derivatives is realized by the AD.

\subsubsection{Loss functions and optimizers}
We consider the minimization problem of the loss function, which is defined as
\begin{equation}
\left\{\bol{\lambda}^\ast, \bol{\gamma}_g^\ast, \bol{\gamma}_\theta^\ast \right\}
 = \underset{\bol{\lambda}, \bol{\gamma}_g, \bol{\gamma}_\theta}{\text{argmin}}
 \left[ \mathcal{J} \left( \bol{\lambda}, \bol{\gamma}_g, \bol{\gamma}_\theta; \mathcal{D}_m, \mathcal{D}_f  \right) \right],  \label{eq-totalLoss}
\end{equation}
where $\left\{\bol{\lambda}^\ast, \bol{\gamma}_g^\ast, \bol{\gamma}_\theta^\ast \right\}$ denote the optimal parameter set.
$\mathcal{D}_f$ is a set of $N_f$ randomly sampled residual points
\begin{equation}
    \mathcal{D}_f = \left\{ \bol{x}_i^f, t_i^f \right\}_{i=1}^{N_f}, 
\end{equation}
and $\mathcal{D}_m$ is the dataset, which contains the spatiotemporal observation data:
\begin{equation}
    \mathcal{D}_m = \left\{ \bol{x}_i^m, t_i^m, \bol{\psi}_i^m \right\}_{i=1}^{N_m}, \label{eq-dataLoss}
\end{equation}
where $\bol{\psi}^m = \{ u, v, w, \phi \}$ is the measured field data.
Note that the pressure $p$ is not included in $\bol{\psi}^m$. In the practical observation of the flow field, such as Particle Image Velocimetry (PIV), obtaining the velocity and pressure at the same time seems to be extremely difficult. For this reason, only the velocity field is assumed to be measured, and pressure is not used in the evaluation of data loss. To be clear, the value of the pressure gradient is inferred by the PINN, which is substituted in the governing equation for the evaluation of physics loss.

The total loss function \cref{eq-totalLoss} is composed of three sub-loss functions defined as 
\begin{subequations}   
\begin{align}
\mathcal{J} \left( \bol{\lambda}, \bol{\gamma}_g, \bol{\gamma}_\theta; \mathcal{D}_m, \mathcal{D}_f  \right) &= \mathcal{J}_p \left( \bol{\lambda}, \bol{\gamma}_g, \bol{\gamma}_\theta; \mathcal{D}_f  \right)
 + \mathcal{J}_d \left( \bol{\lambda}; \mathcal{D}_m  \right)
 + \mathcal{J}_r \left( \bol{\lambda}, \bol{\gamma}_g, \bol{\gamma}_\theta; \mathcal{D}_f  \right), \\
\mathcal{J}_p \left( \bol{\lambda}, \bol{\gamma}_g, \bol{\gamma}_\theta; \mathcal{D}_f  \right)
 &= \sum_{j} \frac{1}{N_f} 
 \left\| \mathcal{F}_j \left[ \bol{\psi}^f_\lambda \paren{\bol{x}^f, t^f}, \bol{\gamma}_g, \bol{\gamma}_\theta \right] \right\|_2^2, \\
\mathcal{J}_d \left( \bol{\lambda}; \mathcal{D}_m  \right)
 &=  \frac{1}{N_m} \left\| \bol{\psi}^f_\lambda \paren{\bol{x}^m, t^m} - \bol{\psi}^m \right\|_2^2, \label{eq-JdDef} \\
\mathcal{J}_r\left( \bol{\lambda}, \bol{\gamma}_g, \bol{\gamma}_\theta; \mathcal{D}_f  \right)
 &= \beta_{\ell_2}  \mathcal{J}_{\ell_2}\paren{ \bol{\lambda}}
 + \beta_g      \mathcal{J}_g \paren{ \bol{\gamma}_g; \mathcal{D}_f  }
 + \beta_\theta \mathcal{J}_\theta \paren{ \bol{\gamma}_\theta; \mathcal{D}_f  }
 + \beta_h      \mathcal{J}_h,  \label{eq-regularizations}
\end{align} 
\end{subequations}
where $\mathcal{J}_p$ is the physics loss, which is defined as the mean-squared error (MSE) of the residual of GEs evaluated on residual points $\mathcal{D}_f$.
$\mathcal{J}_d$ is the data loss, which is the MSE between observed data $\bol{\psi}^m$ and solution fields $\bol{\psi}^f_\lambda$ predicted by the PINN on data points $\{\bol{x}^m,t^m \}$.
$\mathcal{J}_r$ is the regularization loss, which consists of $\ell_2$ regularization $\mathcal{J}_{\ell_2}$, scalings for FreeEnergyNet $\mathcal{J}_g$ and DissipationNet $\mathcal{J}_\theta$, and pressure reference $\mathcal{J}_h$.
$\beta_{\ell_2}, \beta_g, \beta_\theta, \beta_h$ are the weighting factor for the corresponding regularizations.
These regularizations may not be used depending on the problems.

The optimization of \cref{eq-totalLoss} is executed by two optimizers.
The first 1000 epochs are trained with the Adam optimizer, then it switches to the Self-Scaled Broyden (SSBroyden) method.
The SSBroyden method is the 2nd-order quasi-Newton method that has been reported as one of the best optimizers for training PINNs \cite{Kiyani2025, jnini2026curvature}.
For better generalization, residual points are resampled in prescribed intervals with the residual-based adaptive distribution (RAD) \cite{Wu2023}.

\subsection{Implementation and hardware}
The aforementioned methodologies are implemented on \texttt{Python-3.10} code with libraries \texttt{tensorflow-2.16}, \texttt{keras-3.7} and \texttt{scipy-1.16}.
Regarding the SSBroyden optimizer, the implementation by \citet{Kiyani2025} was used.
The code was executed with FP64 floating-point precision on the NVIDIA GH200 installed on Supermicro ARS-111GL-DNHR-LCC and FUJITSU Server PRIMERGY CX2550 M7 (Miyabi) at the Joint Center for Advanced High Performance Computing (JCAHPC), Japan.
Part of the computation was executed on Oscar at the Center for Computation and Visualization (CCV) in Brown University, USA.

\section{Validation}\label{sec-validation}

\begin{table}[bp]
\caption{\label{tab-validation_list} Cases for validation. }
\small
\rowcolors{2}{tb1}{white}
\begin{tabularx}{\textwidth}{lcccccC}
\toprule
 \rowcolor{tbH} \head{Equations} & \head{Dim.} & \head{Steady/Unsteady} & \head{Term discovered} & \head{$g$}  & \head{$\theta$}  & \head{Diff. in data A/B} \\
\midrule
Burgers              & 1D & Unsteady & $\eta u_{xx}$   & --     & $\circ$  & Initial conditions      \\
Kuramoto--Sivashinsky & 1D & Unsteady & $\beta \phi_{xx} + \gamma \phi_{xxxx}$ & $\circ$ & -- & Time regimes \\
Navier--Stokes        & 2D & Steady   & $\nabla^2 \bol{u}/\Rey$      & --      & $\circ$    & Flow types  \\
Navier--Stokes        & 2D & Steady   & $\nabla\cdot\paren{\hat{\eta}\paren{\dot{\gamma}}\bol{D}}/\Rey$     & --      & $\circ$   & Reynolds numbers    \\
\bottomrule
\end{tabularx}
\end{table}

\subsection{Overview}
We demonstrate data-driven discovery of the physics model on several problems to verify our proposed \emph{GIMLET}.
Points of validation are to check
1) whether the free energy and/or dissipation functions are properly discovered, and
2) whether the trained physical model can be transferred to other datasets.
Selected problems for validations are summarized in \Cref{tab-validation_list}.
For each problem, two datasets (A and B) are prepared using conventional CFD solvers.
First, all the neural networks in the \emph{GIMLET} are trained for the dataset A.
Since the true form of the solutions is known, 
the trained free energy function $g\paren{\phi}$ and/or the dissipation function $\theta\paren{\bol{u}}$ are compared with the true function forms.
To assess generalizability, the trained $g\paren{\phi}$ and/or $\theta\paren{\bol{u}}$ are fixed, and only the PINN is trained for dataset B.

As mentioned in \cref{sec-method}, the present \emph{GIMLET} is designed to discover time-independent constitutive laws only. Even for the unsteady Navier--Stokes equations for Newtonian and non-Newtonian fluids, the dissipation functions $\theta$ are time-independent.
Thus, in this study, only steady problems are validated for the Navier--Stokes equations.
The discovery of time-dependent constitutive laws, which involve viscoelastic fluids or turbulent closures, remains as future work.
The network sizes for the showcases were determined from a hyperparameter survey, which is described in \ref{app-params}.
Regarding the random initialization of neural networks, it was confirmed that \emph{GIMLET} can consistently discover the model for different runs with different random seeds.

\subsection{Burgers equation} \label{sec-Burgers}

\subsubsection*{Problem setup}
We first consider a one-dimensional viscous Burgers' equation expressed as
\begin{equation}
\pdif{u}{t} + u\pdif{u}{x} - \eta\plap{u}{x} = 0, \label{eq-Burgers1}
\end{equation}
in a periodic spatial domain $x\in [-1,1]$ and time range $t \in [0,1]$ with diffusion coefficient $\eta = 0.01 / \pi$.
The equivalent form of \cref{eq-Burgers1} can be written as
\begin{equation}
\pdif{u}{t} + u\pdif{u}{x} + \frac{\delta\Theta}{\delta u} = 0, \label{eq-Burgers2}
\end{equation}
using the dissipation function 
\begin{equation}
\theta\paren{\pdif{u}{x}} = \frac{\eta}{2} \paren{\pdif{u}{x}}^2,
\end{equation}
which is a target to be discovered by the proposed method.

\subsubsection*{Datasets}
Two datasets are prepared with different initial conditions:
\begin{subequations}
\begin{alignat}{2}
\text{Dataset A} \qquad & u\paren{x, t=0} = - \sin\paren{\pi x}, \\
\text{Dataset B} \qquad & u\paren{x, t=0} = \exp\paren{-\frac{\paren{x-x_0}^2}{2\sigma^2}},
\end{alignat}
\end{subequations}
with $x_0=0.2$ and $\sigma = 0.1$.
The datasets are calculated by the Fourier spectral method with $N_x = 512$ modes, and time integration by the Runge--Kutta 4th order scheme with time interval $\Delta t = \num{1e-5}$.
The number of stored time points is $N_t = 101$, with an output interval $\Delta t_\text{out} = \num{1e-2}$.

\subsubsection*{Network design and optimization}
For this problem, the FreeEnergyNet is not used, and the weights for the PINN and the DissipationNet are trained.
For the PINN, 8 fully-connected dense hidden layers with 20 neurons each are used with $\tanh$ activation function.
For the DissipationNet, 4 hidden layers with 10 neurons each are used with softplus and sigmoid activation functions. 
After the \num{1000} epochs of the Adam optimizer stage, the SSBroyden optimizer is executed for \num{10000} epochs.
For the physics loss, $N_f = \num{5000}$ residual points are used for the training.
For better generalization, residual points are resampled in every \num{1000} by the RAD method.
For the data loss, the $N_d = \num{5000}$ data points are randomly selected from the total $N_x\times N_t = 512 \times 101$ generated dataset.
From the remaining data points, which are not selected for data loss evaluation, $N_e = \num{1250}$ points are selected for test evaluation.
For the regularization term $\mathcal{J}_r$ in the loss function, $\ell_2$ regularization for the PINN is employed with coefficient $\beta_{\ell_2} = \num{1e-11}$.
Other regularization terms in \cref{eq-regularizations} are not used and $\beta_g = \beta_\theta = \beta_h = 0$.

\begin{figure}[tp]
    \centering
    \includegraphics[width=\textwidth]{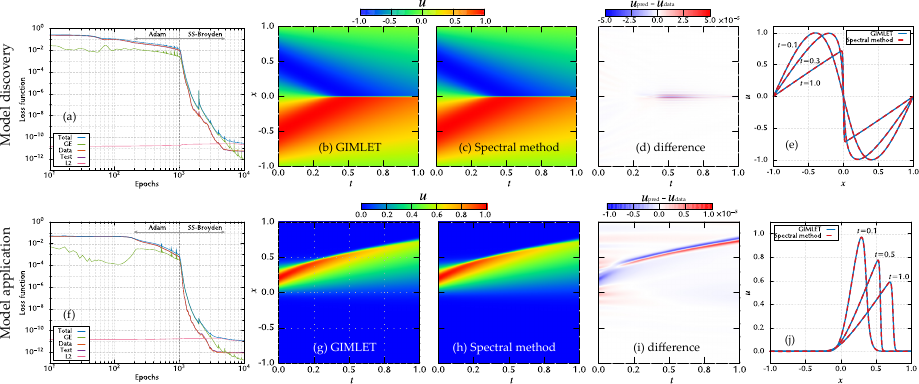}
    \caption{Results of model discovery and its application for Burgers equation. 
    Upper panels are for model discovery, and lower panels are for model application.
    (a,f): history of loss function, where the total value, $\mathcal{J}_p$ (GE), $\mathcal{J}_d$ (Data), and the test loss (Test) are separately plotted.
    (b,g): Spatio-temporal distribution of solution calculated by \emph{GIMLET}.
    (c,h): Solutions by the spectral method.
    (d,i): Differences between predictions and data.
    (e,j): Selected snapshots.}
    \label{fig-Burgers1}
\end{figure}
\begin{figure}[tp]
    \centering
    \includegraphics[width=\textwidth]{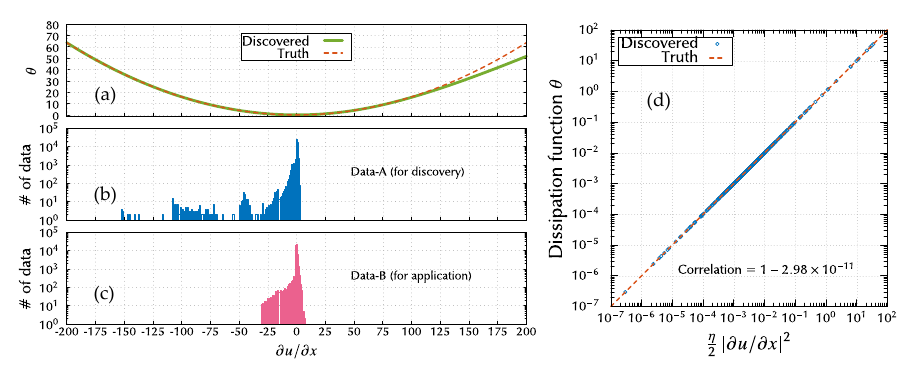} 
    \caption{(a) Dissipation function $\theta(u_x)$, where the green solid line indicates the function discovered by the \emph{GIMLET} and the orange dashed line the true function.
    (b,c) The histograms of $u_x$ in the datasets used for discovery and application, respectively. 
    (d) Relation between trained $\theta$ and $\nicefrac{\eta}{2} \paren{u_x}^2$, where the blue circles are values in the dataset A.}
    \label{fig-Burgers2}
\end{figure}

\subsubsection*{Results}
Results for model discovery and generalization check are presented in \cref{fig-Burgers1,fig-Burgers2}, which show satisfactory performance of the proposed method.
The upper panels in \cref{fig-Burgers1} show the results of the model discovery stage, where DissipationNet is trained.
It can be clearly seen that the loss function (\cref{fig-Burgers1}a) converged lower than \num{1e-10}, and the predicted spatiotemporal solution (\cref{fig-Burgers1}b) showed good agreement (\cref{fig-Burgers1}d,e) with the solution calculated by the spectral method (\cref{fig-Burgers1}c).
The maximum absolute errors between the predicted and reference solutions are \num{5e-5} for model discovery (\cref{fig-Burgers1}d) and \num{1e-5} for model application (\cref{fig-Burgers1}i).
The discovered dissipation function $\theta\paren{u_x}$ is shown in \cref{fig-Burgers2} with histograms of the $u_x$ involved in the datasets.
In the dataset A, the values of the gradient $u_x$ distribute approximately in $[-152, 3.2]$ with a large absolute lower bound, due to the steep gradient at $x=0$.
The discovered dissipation function $\theta$ agreed almost completely with the true function as shown in \cref{fig-Burgers2}a, even for the outside of the data range (\cref{fig-Burgers2}b).
This result leads to good generalizability, as shown below.
More quantitative validation is shown in \cref{fig-Burgers2}d, where the trained $\theta$ is plotted as a function of $\eta/2 \paren{u_x}^2$ for dataset A.
The correlation coefficient is $1 - \num{2.98e-11}$, whose deviation from unity is sufficiently small.

For the generalization check, the trained DissipationNet is fixed and is applied to the dataset B, which shows considerably different behavior from the dataset A.
It is confirmed that the loss function converged sufficiently (\cref{fig-Burgers1}f), the predicted solution (\cref{fig-Burgers1}g) showed good agreement (\cref{fig-Burgers1}i,j) with the solution calculated by the spectral method (\cref{fig-Burgers1}h).
The histogram of $u_x$ in dataset B is shown in \cref{fig-Burgers2}c, and ranged in $[-30, 7.6]$.
Although the upper bound on $u_x$ in dataset B is larger than that in dataset A, the PDE with the discovered $\theta\paren{u_x}$ almost completely captures dataset B.

\subsection{Kuramoto--Sivashinsky equation} \label{sec-KS}

\subsubsection*{Problem setup}
For a PDE involving higher-order derivatives, we consider the one-dimensional Kuramoto--Sivashinsky equation expressed by
\begin{equation}
 \pdif{\phi}{t} + \alpha \phi\pdif{\phi}{x} + \beta \plap{\phi}{x} + \gamma \frac{\partial^4 \phi}{\partial x^4} = 0, \label{eq-KS1}
\end{equation}
with $\alpha = 100 / 16$, $\beta = 100 / (16^2)$, $\gamma = 100 / (16^4)$.
The solution domain is defined as $\paren{t,x} \in [0,1] \times [0, 2\pi]$.
We assume that the time derivative term $\phi_t$ and the convective term $\alpha\phi\phi_x$ are known, and the remaining two terms are to be discovered.
The equivalent form of \cref{eq-KS1} can be written as
\begin{equation}
   \pdif{\phi}{t} + \alpha \phi\pdif{\phi}{x} -  \plap{}{x}\paren{\frac{\delta G}{\delta \phi}} = 0, \label{eq-KS2}
\end{equation}
using the free energy function
\begin{equation}
    g\paren{\phi, \pdif{\phi}{x}} = -\frac{\beta}{2} \phi^2 + \frac{\gamma}{2}\paren{\pdif{\phi}{x}}^2,
\end{equation}
which will be discovered by the proposed method.

\subsubsection*{Datasets}
The datasets are drawn from the open dataset provided by \citet{Wang2024}, which is a numerical solution of \cref{eq-KS1} for the initial condition
\begin{equation}
    \phi(x,t=0) = \cos\left( x \right) \left( 1 + \sin\left(x \right) \right),
\end{equation}
and calculated by a Fourier spectral method with $512$ modes and a 4th-order Runge--Kutta time-stepping scheme with time-step size \num{1e-5}.
The total stored data size is $N_t \times N_x = 501 \times 512$.
We prepared two datasets by subsampling for two different time regimes: $t \in [0.45, 0.5]$ for dataset A and $t \in [0.7, 0.8]$ for dataset B.
In the time regime for the dataset A, the spatiotemporal variation of $\phi$ is relatively simple, whereas it becomes complex in the later time regime for the dataset B.

\subsubsection*{Network design and optimization}
For this problem, the DissipationNet is not used, and the FreeEnergyNet is composed of two INNs for single-valued functions as
\begin{equation}
    \frac{\delta G}{\delta \phi}\paren{\phi, \phi_x} 
    = \pdif{g}{\phi}\paren{\phi} - \pdif{}{x}\paren{\pdif{g}{\phi_x}}\paren{\phi_x}
    = \Gamma_g \left[  g'_0\paren{\phi} - \pdif{}{x} \paren{ g'_1\paren{\phi_x} }  \right],
\end{equation}
where both $g'_0$ and $g'_1$ are designed as 4 fully-connected dense hidden layers with 10 neurons each.
For the PINN, 5 fully-connected dense hidden layers with 30 neurons each are used with $\tanh$ activation function.
To improve the condition of the problem, the output of FreeEnergyNet is scaled as $\mathcal{O}\paren{1}$ with an additional loss function:
\begin{equation}
    \mathcal{J}_g\paren{\bol{\gamma}_g; \mathcal{D}_p} = \paren{1 - \frac{1}{N_f} \sum_{i=1}^{N_f} 
    \left| g'_0\paren{\phi\paren{x^p_i,t^p_i}} - \pdif{g'_1}{x}\paren{\phi_x\paren{x^p_i,t^p_i}} \right| }^2
\end{equation}
and the coefficient $\Gamma_g$ is multiplied to the functional derivative term as in \cref{eq-KS2}, and $\Gamma_g$ is treated as an additional trainable parameter in the optimization.
For the regularization terms $\mathcal{J}_r$ in \cref{eq-regularizations}, $\ell_2$ regularization for the PINN is employed with coefficient $\beta_{\ell_2} = \num{1e-11}$.
The weight the regularization $\mathcal{J}_g$ is selected as $\beta_g = 1$, and other regularizations are not used ($\beta_\theta = \beta_h = 0$).
After the \num{1000} epochs of the Adam optimizer stage, the SSBroyden optimizer is executed for \num{200000} epochs with FP64 double precision.
For the physics loss, $N_f = \num{30000}$ residual points are used for the training with the RAD resampling in every \num{1000} epochs.
For the data loss, the $N_d = \num{5000}$ data points are randomly selected from the dataset subsampled for each time regime, which is taken from the total stored data of $N_t \times N_x = 501 \times 512$.
From the remaining data points, which are not selected for data loss evaluation, $N_e = \num{1250}$ points are selected for test evaluation.

\subsubsection*{Results}

\begin{figure}[tbp]
    \centering\includegraphics[width=\textwidth]{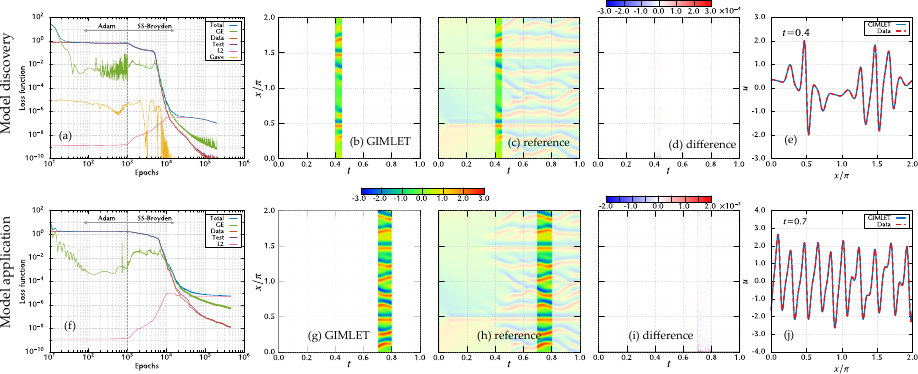}
    \caption{Results of model training for Kuramoto--Sivashinsky equation. 
    Upper panels are for model discovery, and lower panels are for model application.
    (a,f): History of loss function, where the total value, $\mathcal{J}_p$ (GE), $\mathcal{J}_d$ (Data), and the test loss (Test) are separately plotted.
    (b,g): Spatio-temporal distribution of solution calculated by \emph{GIMLET}.
    (c,h): Solutions by the spectral method, where the time regimes shown in half-transparent are not used for training.
    (d,i): Differences between predictions and data.
    (e,j): Selected snapshots.}
    \label{fig-KS1}
\end{figure}

\begin{figure}[tbp]
    \centering
    \includegraphics[width=\textwidth]{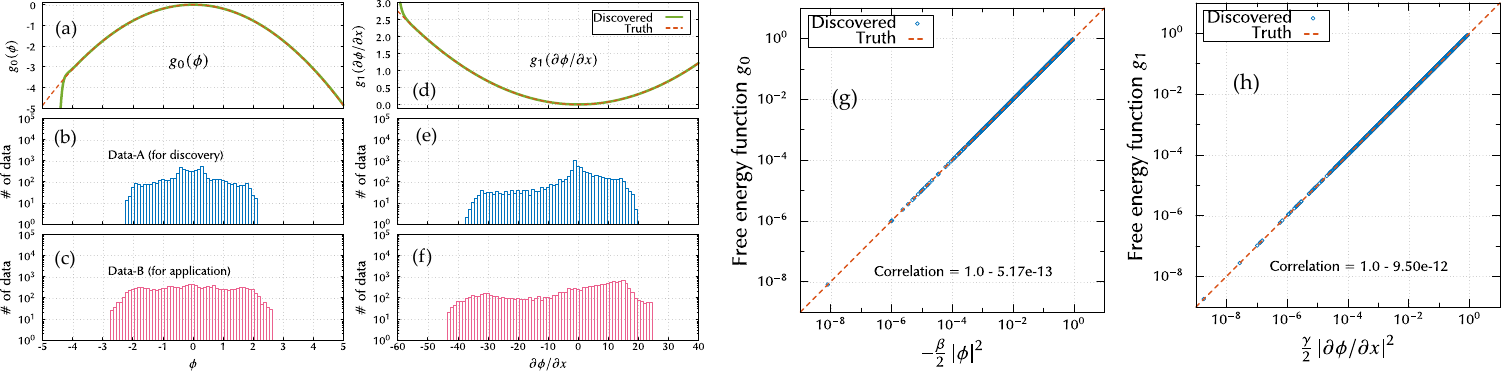} 
    \caption{(a,d): Free energy functions $g_0\paren{\phi}$ and $g_1\paren{\phi_x}$, where the green solid lines indicate the functions discovered by the \emph{GIMLET} and the orange dashed lines the true functions.
    (b,e): The histograms of $\phi$ and $\phi_x$ in the dataset A used for discovery. 
    (c,f): The histograms in the dataset B for the generalization check.
    (g): Relation between trained $g_0$ and $-\beta/2 \paren{\phi}^2$, where the blue circles are values in the dataset A.
    (h): Relation between $g_1$ and $\gamma/2 \paren{\phi_x}^2$.}
    \label{fig-KS2}
\end{figure}

Results for model discovery and generalization are presented in \cref{fig-KS1,fig-KS2}.
The upper panels in \cref{fig-KS1} show the results of the model discovery stage, where FreeEnergyNet is trained.
It can be clearly seen that the loss function (\cref{fig-KS1}a) converged lower than \num{1e-9}, and the predicted spatiotemporal solution (\cref{fig-KS1}b) showed good agreement (\cref{fig-KS1}d,e) with the reference solution (\cref{fig-KS1}c).
The maximum absolute errors between the predicted and reference solutions are \num{3e-4} for model discovery (\cref{fig-KS1}d) and \num{2e-3} for model application (\cref{fig-KS1}i).
The discovered free energy functions $g_0\paren{\phi}$ and $g_1\paren{\phi_x}$ are shown in \cref{fig-KS2} with histograms of $\phi$ and $\phi_x$ involved in the datasets.
In the dataset A, the values $\phi$ and $\phi_x$ distribute approximately in $[-2.3, 2.2]$ and $[-38, 20]$, respectively.
The discovered free energy functions $g_0$ and $g_1$ agreed almost completely with the true function as shown in \cref{fig-KS2}a,d, even for the outside of the data range (\cref{fig-KS2}b,e).
\Cref{fig-KS2}g shows the trained $g_0$ plotted as a function of $-\beta/2\paren{\phi}^2$ for dataset A, whereas \cref{fig-KS2}h shows $g_1$ as a function of $\gamma/2\paren{\phi_x}^2$. 
The correlation coefficients for $g_0$ and $g_1$ are $1 - \num{5.17e-13}$ and $1 - \num{9.50e-12}$, whose deviations from unity are both sufficiently small.

For the generalization check, the trained FreeEnergyNet is fixed and is applied to the dataset B, which shows more complex behavior than the dataset A.
It is confirmed that the loss function converged sufficiently (\cref{fig-KS1}f), and the predicted solution (\cref{fig-KS1}g) showed good agreement (\cref{fig-KS1}i,j) with the solution calculated by the spectral method (\cref{fig-KS1}h).
The data ranges in $\phi$ and $\phi_x$ in dataset B, which are shown in \cref{fig-KS2}(c,f), are wider than those in dataset A.
Nevertheless, the PDE with the discovered $g_0$ and $g_1$ almost completely captures dataset B, which indicates the good generalizability of the trained model.

\subsection{Constitutive model for Newtonian fluids} \label{sec-NS}

\subsubsection*{Problem setup}
For a multi-dimensional system, we next consider the two-dimensional steady fluid flow of a Newtonian fluid, which is governed by the following continuity and Navier--Stokes equations:
\begin{subequations}
\label{eq-NS1}
\begin{gather}
    \nabla\cdot\bol{u} = 0,  \\
    \bol{u}\cdot \nabla \bol{u} = -\nabla p + \frac{1}{\Rey} \nabla^2 \bol{u}, \label{eq-NS1b}
\end{gather}
\end{subequations}
where $\bol{u}$ and $p$ are velocity and pressure, respectively.
The variables are nondimensionalized, and $\Rey= \rho U d / \mu_\text{f}$ is the Reynolds number, where $U$, $d$, $\rho$, and $\mu_\text{f}$ are the characteristic velocity, characteristic length, density, and dynamic viscosity, respectively. Note that $\mu_\text{f}$ here is the viscosity, not the chemical potential $\mu$ of \cref{sec-method}.
Assuming the convective term and the pressure gradient are known, the viscous term is unknown and needs to be discovered.
The equivalent form of \cref{eq-NS1b} can be written using the dissipation function as
\begin{subequations}
\label{eq-NS2}
\begin{gather}
    \bol{u}\cdot \nabla \bol{u} 
    = -\nabla p - \Gamma_\Rey \frac{\delta \Theta}{\delta \bol{u}}, \label{eq-NS2a} \\
    \theta\paren{\nabla\bol{u}} = \frac{1}{2\Rey} \paren{\nabla\bol{u}}^2, \label{eq-NS2b}
\end{gather}
\end{subequations}
where $\theta$ is the target to be discovered.
$\Gamma_\Rey$ is a linear coefficient, which is introduced to handle the difference in $\Rey$ between datasets.

\subsubsection*{Datasets}
We generated two quite different datasets for this problem.
For the dataset A, which is used for the model discovery, the lid-driven cavity flow of $\Rey=400$ is considered.
The solution is calculated by the finite volume method using \texttt{icoFoam} implemented on \texttt{OpenFOAM-v2412}.
The spatial domain $\paren{x,y} \in [0,1] \times [0,1]$ is uniformly discretized with $200\times 200$ square cells.
Regarding the boundary conditions, the uniform velocity at the top wall causes a singularity at the top-right and top-left corners.
To avoid this singularity, the following regularization is applied at the moving wall boundary at $y=1$:
\begin{equation}
\label{eq-Utop}
u\paren{x,y=1} = \begin{cases}
    \frac{1}{4} \left(1 - \cos\paren{10\pi x}\right)^2, & x < 0.1 \\
    1, & 0.1 \leq x \leq 0.9 \\
    \frac{1}{4} \left(1 - \cos\paren{10\pi x}\right)^2, & 0.9 < x
    \end{cases}    
\end{equation}

For the dataset B, which is used for the generalizability check, the flow past a cylinder of $\Rey=20$ is considered.
The solution is calculated by the spectral/hp element method~\cite{Karniadakis2005} as implemented in \texttt{Nektar++}. 
For the data generation, the spatial domain of $\paren{x,y} \in [-7.5, 22.5] \times [-10,10]$ around the cylinder of diameter $d=1$ is discretized into \num{34368} quadrilateral elements.
A uniform inflow boundary condition $(u=1,v=0)$ is applied at the inlet $x=-7.5$ boundary.
The outflow boundary $(\partial\bol{u}/\partial\bol{n}=0, p=0)$ is imposed on $x=22.5$, where $\bol{n}$ is the normal vector at the outlet boundary.
On the lateral boundaries ($y=\pm 10)$, symmetric boundary conditions are applied.
After calculating the solution, the sub-region of $\paren{x,y} \in [-2,2] \times [-2,2]$ near the cylinder is extracted and used as the dataset.

Note, as described in \cref{sec-method}, the pressure calculated by the CFD is not used in the evaluation of data losses. 
The data loss is evaluated by only the error in the predicted and measured velocity fields.

\subsubsection*{Network design and optimization}
For this problem, the FreeEnergyNet is not used, and the dissipation function is defined as a four-variable function as $\theta\paren{\nabla\bol{u}} = \theta\paren{u_x, u_y, v_x, v_y}$, and expressed by a neural network of 4 fully-connected dense hidden layers with 10 neurons each.
For the PINN, 8 fully-connected dense hidden layers with 20 neurons each are used with $\tanh$ activation function.
The same structure is used for both model discovery and generalizability check.

In the incompressible Navier--Stokes equation with all Neumann boundary conditions, the reference value of the pressure is required, because only the gradient of the pressure appears in the governing equation.
For this reason, the following loss function is introduced as an additional regularization term to impose the pressure reference:
\begin{equation}
    \mathcal{J}_h = \left\{ p\paren{x_p, y_p} \right\}^2, \label{eq-pRef}
\end{equation}
where $\paren{x_p,y_p}$ is a reference point, which is selected as $\paren{x_p,y_p} = \paren{0.5, 1}$ for the lid-driven cavity problem, and $\paren{x_p,y_p} = \paren{2, 0}$ for the problem of flow past a cylinder.

The Reynolds number for the dataset B ($\Rey=20$) is different from that for the dataset A ($\Rey=400$), so the dissipation functions for the two datasets must be different by the ratio $\Rey_A/\Rey_B$ while the function form of $\theta$ is common.
To deal with this difference in $\Rey$, the coefficient $\Gamma_\Rey$ is multiplied by the dissipation function term in the governing equation \cref{eq-NS2a}.
In the discovery stage, this coefficient is fixed as $\Gamma_\Rey = 1$.
Then, for the generalization check stage, we let $\Gamma_\Rey$ be an additional training parameter in the optimization.

After the \num{1000} epochs of the Adam optimizer stage, the SSBroyden optimizer is executed for \num{100000} epochs with FP64 double precision.
For the physics loss, $N_f = \num{20000}$ residual points are used for the training with the RAD resampling in every \num{1000} epochs.
For the data loss, $N_d = \num{10000}$ data points are randomly selected from the original datasets.
From the remaining data points, which are not selected for data loss evaluation, $N_e = \num{2500}$ points are selected for test evaluation.
For the regularization term $\mathcal{J}_r$ in the loss function, $\ell_2$ regularization for the PINN is employed with coefficient $\beta_{\ell_2} = \num{1e-11}$.
The reference point of the pressure is introduced as $\paren{x_p,y_p} = \paren{0.5, 1}$ with the weight $\beta_h = 1$.
Other regularization terms in \cref{eq-regularizations} are not used ($\beta_g = \beta_\theta=0$).

\subsubsection*{Results}

\begin{figure}[tbp]
    \centering
    \includegraphics[width=\textwidth]{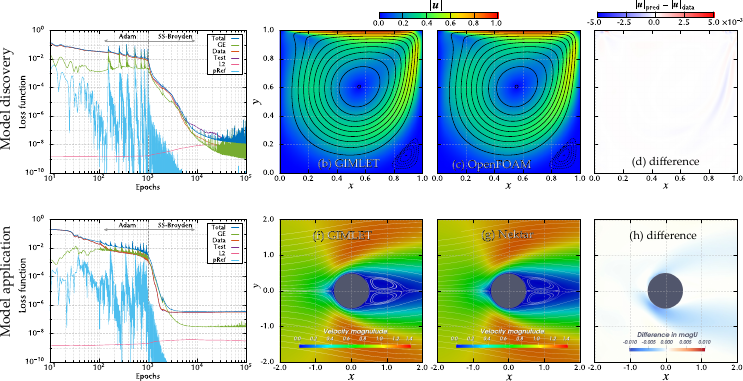}
    \caption{Results of model training for dissipation function of Newtonian fluids.
    The upper panels are for model discovery, and the lower panels are for model application.
    (a,e): History of loss function, where the total value, $\mathcal{J}_p$ (GE), $\mathcal{J}_d$ (Data), test loss (Test), $\ell_2$ loss (L2), and loss for pressure reference (pRef) are separately plotted.
    (b,f): Flow field calculated by \emph{GIMLET}. Streamlines (black solid lines) and velocity magnitude (color contour).
    (c): Reference flow field calculated by \texttt{OpenFOAM}.
    (g): Reference flow field calculated by \texttt{Nektar++}.
    (d,h): Differences of velocity magnitude between \emph{GIMLET} and reference data.}
    \label{fig-NS1}
\end{figure}

\begin{figure}[tbp]
    \centering
    \begin{tabular}{cc}
    \includegraphics[width=0.38\textwidth]{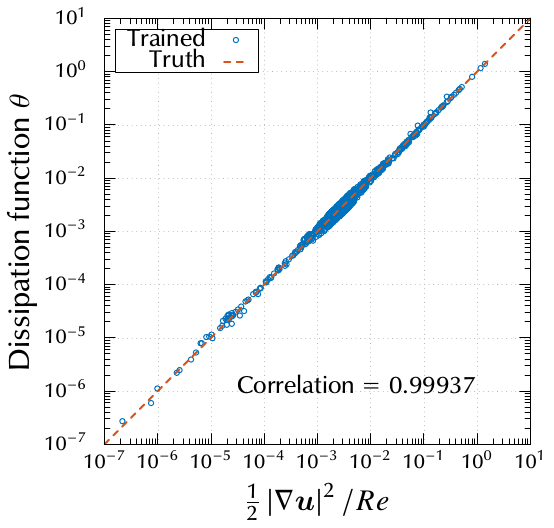} & 
    \includegraphics[width=0.58\textwidth]{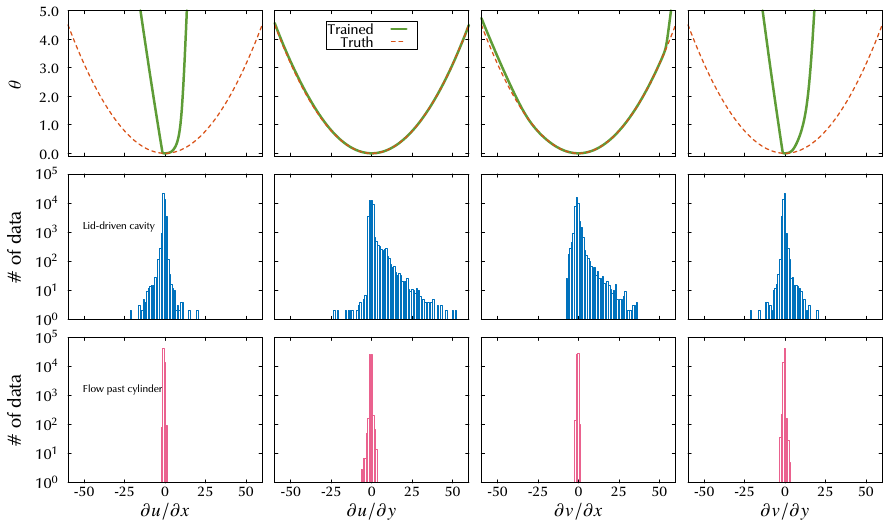} \\
    (a) & (b)
    \end{tabular}   
    \caption{(a) Comparison between true and trained dissipation functions plotted for the dataset A.
    (b) Dissipation function projected to one dimension (upper) and data distributions corresponding to the projected axes (middle and lower).}
    \label{fig-NS2}
\end{figure}

Results for model discovery and generalization check are presented in \cref{fig-NS1,fig-NS2}.
Upper panels in \cref{fig-NS1} show the results of the model discovery stage, where DissipationNet is trained.
It can be clearly seen that the loss function (\cref{fig-NS1}a) converged down to the order of \num{1e-8}, and the predicted flow field (\cref{fig-NS1}b) showed good agreement (\cref{fig-NS1}d) with the reference solution (\cref{fig-NS1}c).
The maximum absolute errors between the predicted and reference solutions are \num{5e-3} for model discovery (\cref{fig-NS1}d) and \num{1e-2} for model application (\cref{fig-NS1}h).

The discovered dissipation function $\theta$ is shown in \cref{fig-NS2} with histograms of $\nabla\bol{u}$ involved in the datasets.
\Cref{fig-NS2}a shows a comparison between true and trained dissipation functions plotted for dataset A.
The correlation coefficient is evaluated as \num{0.99937}, which means that the DissipationNet is almost completely fit to the given training data.
To see the range outside the training data, the four projected functions of $\theta$ are plotted in the top panels of \cref{fig-NS2}b.
Note that the trained function is a four-variable function $\theta\paren{u_x,u_y,v_x,v_y}$, and the functions plotted in the figure are projected onto one axis; for instance, the top left panel is $\theta\paren{u_x, 0, 0, 0}$.
It can be seen that $\theta$ is discovered well for the range of velocity gradient $\nabla\bol{u}$ involved in the dataset A.
For the range outside of the dataset, the dissipation function $\theta$ still keeps accuracy for $u_y$ and $v_x$. 
However, for $u_x$ and $v_y$, the applicable range is quite narrow.
The dataset A is the lid-driven cavity flow, where the moving top wall generates strong shear.
This shear appears in $u_y$ and $v_x$ in the coordinate system employed.

For the generalization check, the trained DissipationNet is fixed and is applied to the dataset B, which is a quite different flow type from the dataset A.
It is confirmed that the loss function converged sufficiently (\cref{fig-NS1}e), the predicted solution (\cref{fig-NS1}f) showed good agreement (\cref{fig-NS1}h) with the solution calculated by the spectral/hp element method (\cref{fig-NS1}g).

\begin{figure}[tbp]
    \centering
    \includegraphics[width=0.50\textwidth]{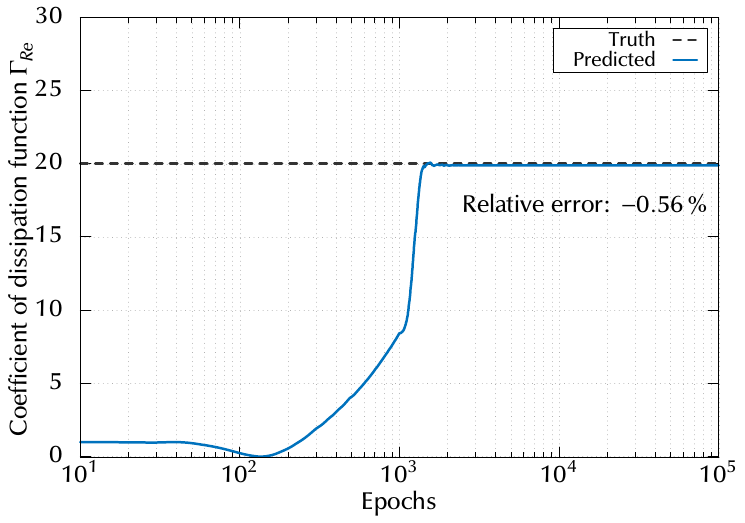}
    \caption{History of coefficient for dissipation function $\Gamma_\Rey$.}
    \label{fig-NS3}
\end{figure}

The history of the coefficient for the dissipation function $\Gamma_\Rey$ is shown in \cref{fig-NS3}.
The dissipation function $\theta$ trained for $\Rey_A=400$ is applied to the flow at $\Rey_B=20$, thus, the ratio of Reynolds numbers is $\Rey_A / \Rey_B = 20$.
From \cref{fig-NS3}, one can observe that $\Gamma_\Rey$, initially set to 1, converged to a value very close to 20, with a relative error of only \SI{0.56}{\percent}.
It can be seen that the data range of $\nabla\bol{u}$ in dataset B (flow past a cylinder) overlaps that of dataset A (lid-driven cavity flow), and this enabled successful generalization across quite different flow types.

\subsubsection*{Generalizability improvement}\label{sec-argument}
\begin{figure}[tbp]
    \centering
    \begin{tabular}{cc}
    \includegraphics[width=0.44\textwidth]{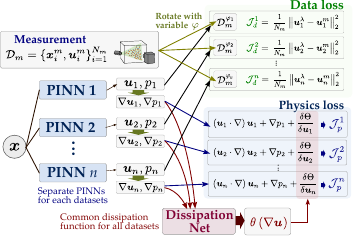} & 
    \includegraphics[width=0.52\textwidth]{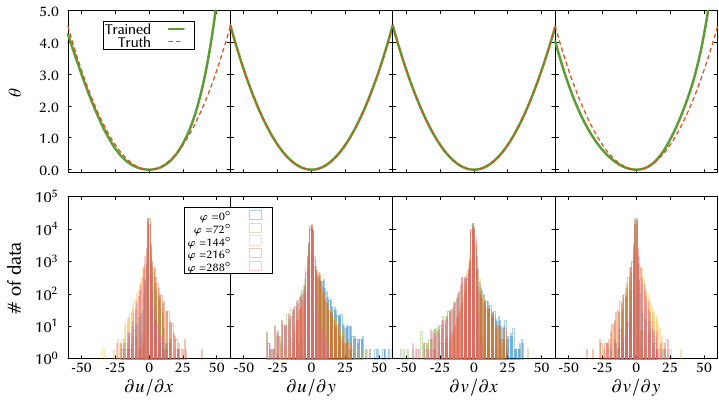} \\
    (a) & (b)
    \end{tabular}   
    \caption{(a) Architecture for multiple rotated datasets. 
    (b) Trained dissipation functions projected to one dimension (upper), and data distributions corresponding to the projected axes (lower).}
    \label{fig-NS4}
\end{figure}

Through the above-mentioned validation, we understood that the range of velocity gradient $\nabla\bol{u}$ in the dataset for model discovery is significant for good generalizability.
By the wider range of $\nabla{\bol{u}}$, better generalizability can be expected.
One possible way to improve the generalizability is data augmentation.
In this study, we demonstrate the effect of data rotations.
\Cref{fig-NS4}a shows the architecture of \emph{GIMLET} for multiple dataset training.
Suppose we have $n$ datasets, and we use $n$ of independent PINNs to predict solutions for each dataset.
A single DissipationNet is used across all datasets, and all PINNs and the DissipationNet are trained simultaneously.
In this study, five datasets are produced by rotating the original dataset A by different angles.
The dissipation function $\theta$ trained in this way is shown in \cref{fig-NS4}b with histograms of datasets.
It can be seen that the error between trained and true $\theta$ is considerably improved compared to the single dataset training (\cref{fig-NS2}b).

\subsection{Constitutive model for non-Newtonian fluids} \label{sec-NonNewton}
\subsubsection*{Problem setup}
For a more complicated and practical problem, we consider the fluid flow of a non-Newtonian fluid.
Assuming a two-dimensional steady state, the governing equations can be written as
\begin{subequations}
\label{eq-nonNewton1}
\begin{gather}
    \nabla\cdot\bol{u} = 0,  \\
    \bol{u}\cdot \nabla \bol{u}     = -\nabla p 
    + \frac{1}{\Rey} \nabla\cdot\paren{\hat{\eta}\paren{\dot{\gamma}} \paren{\nabla\bol{u} + \paren{\nabla\bol{u}}^t}}, \label{eq-nonNewton1b}
\end{gather}
\end{subequations}
where $\bol{u}$ and $p$ are velocity and pressure, respectively, and $\bol{D} \equiv \nabla\bol{u} + \paren{\nabla\bol{u}}^t$ is twice the rate-of-strain tensor.
$\hat{\eta}\paren{\dot\gamma}$ is the nondimensional effective viscosity depending on the scalar measure of the strain-rate tensor
\begin{equation}
    \dot\gamma
    = \left[ 2\paren{\pdif{u}{x}}^2 + 2\paren{\pdif{v}{y}}^2 + \paren{\pdif{u}{y} + \pdif{v}{x}}^2 \right]^{\frac{1}{2}}
    = \paren{\tfrac{1}{2}\, \bol{D} : \bol{D}}^{\frac{1}{2}}.
\end{equation}
For the viscosity function, we apply the Bird--Cross--Carreau--Yasuda model \cite{Cross1965,Carreau1972,Yasuda1979,Bird1986,Gallagher2019}:
\begin{equation}
    \hat{\eta}\paren{\dot\gamma} = \hat{\eta}_\infty + \left( \hat{\eta}_0 - \hat{\eta}_\infty \right)
    \left[ 1 + \paren{\lambda \dot\gamma}^a \right]^{\frac{n-1}{a}},  \label{eq-BCCY}
\end{equation}
with $\hat{\eta}_0 = 1$, $\hat{\eta}_\infty = 0.5$, $a=2$, $\lambda=5$, and $n=0.5$.

Assuming the convective term and the pressure gradient are known, the viscous term is unknown and to be discovered.
The equivalent form of \cref{eq-nonNewton1b} can be written using the dissipation function as
\begin{subequations}
\label{eq-nonNewton2}
\begin{gather}
    \bol{u}\cdot \nabla \bol{u} 
    = -\nabla p - \Gamma_\Rey \frac{\delta \Theta}{\delta \bol{u}}, \label{eq-nonNewton2b} \\
    \theta\paren{\nabla\bol{u}} = \frac{\hat{\eta}}{2\Rey}\paren{\nabla\bol{u}}^2,
\end{gather}
\end{subequations}
where $\theta$ is the target to be discovered.
$\Gamma_\Rey$ is a linear coefficient, which is introduced to handle the difference in $\Rey$ between datasets.

\subsubsection*{Datasets}
The lid-driven cavity flow problem is considered, and two datasets are prepared with different Reynolds numbers: $\Rey_A=400$ for the dataset A (model discovery) and $\Rey_B=100$ for the dataset B (generalization check).
The solutions are calculated by the \texttt{nonNewtonianIcoFoam} implemented on \texttt{OpenFOAM-v2412}.
The spatial domain $\paren{x,y} \in [0,1] \times [0,1]$ is uniformly discretized with $200\times 200$ square cells.
As is the case with Newtonian fluid, regularization of moving wall velocity \cref{eq-Utop} is applied.
Only the velocity fields are used for the training.

\subsubsection*{Network design and optimization}
Similar to the Newtonian fluid problem, the FreeEnergyNet is not used in this problem, and the dissipation function is defined as a four-variable function as $\theta\paren{\nabla\bol{u}} = \theta\paren{u_x, u_y, v_x, v_y}$, and expressed by a neural network of 4 fully-connected dense hidden layers with 10 neurons each.
For the PINN, 8 fully-connected dense hidden layers with 20 neurons each are used with $\tanh$ activation function.
The same structure is used for both model discovery and generalizability check.
After the \num{1000} epochs of the Adam optimizer stage, the SSBroyden optimizer is executed for \num{100000} epochs with FP64 double precision.
For the physics loss, $N_f = \num{30000}$ residual points are used for the training with the RAD resampling at every 1000 epochs.
For the data loss,  $N_d = \num{30000}$ data points are randomly selected from the original datasets.
From the remaining data points, which are not selected for data loss evaluation, $N_e = \num{7500}$ points are selected for test evaluation.
For the regularization term $\mathcal{J}_r$ in the loss function, $\ell_2$ regularization for the PINN is employed with coefficient $\beta_{\ell_2} = \num{1e-11}$.
The reference point of the pressure is introduced as $\paren{x_p,y_p} = \paren{0.5, 1}$ with the weight $\beta_h = 1$.
Other regularization terms in \cref{eq-regularizations} are not used ($\beta_g = \beta_\theta=0$).
Similar to the Newtonian fluid problem, the coefficient $\Gamma_\Rey$ is multiplied by the dissipation function term in the governing equation \cref{eq-nonNewton2}.
In the discovery stage, this coefficient is fixed as $\Gamma_\Rey = 1$.
Then, for the generalization check stage, we let $\Gamma_\Rey$ be an additional training parameter in the optimization.

\begin{figure}[tbp]
    \centering
    \includegraphics[width=\textwidth]{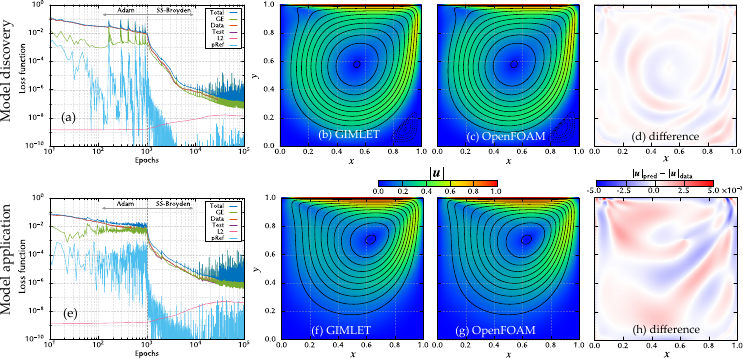}
    \caption{Results of model training for dissipation function of non-Newtonian fluids. 
    The upper panels are for model discovery, and the lower panels are for model application.
    (a,e): History of loss function, where the total value, $\mathcal{J}_p$ (GE), $\mathcal{J}_d$ (Data), test loss (Test), $\ell_2$ loss (L2), and loss for pressure reference (pRef) are separately plotted.
    (b,f): Flow field calculated by \emph{GIMLET}. Streamlines (black solid lines) and velocity magnitude (color contour).
    (c,g): Reference flow field calculated by OpenFOAM.
    (d,h): Differences of velocity magnitude between \emph{GIMLET} and reference data.}
    \label{fig-nonNewton1}
\end{figure}

\begin{figure}[tbp]
    \centering
    \begin{tabular}{cc}
    \includegraphics[width=0.38\textwidth]{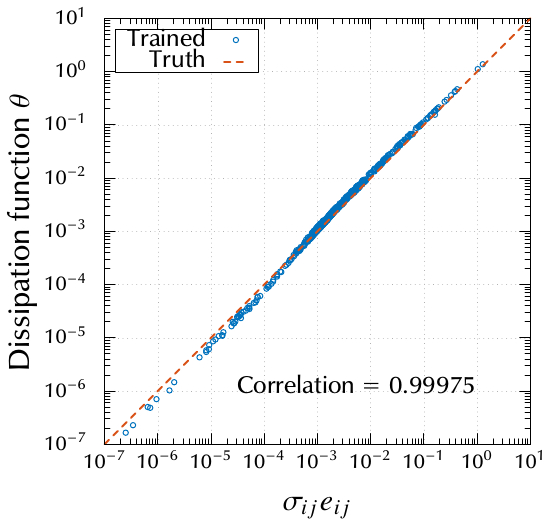} & 
    \includegraphics[width=0.58\textwidth]{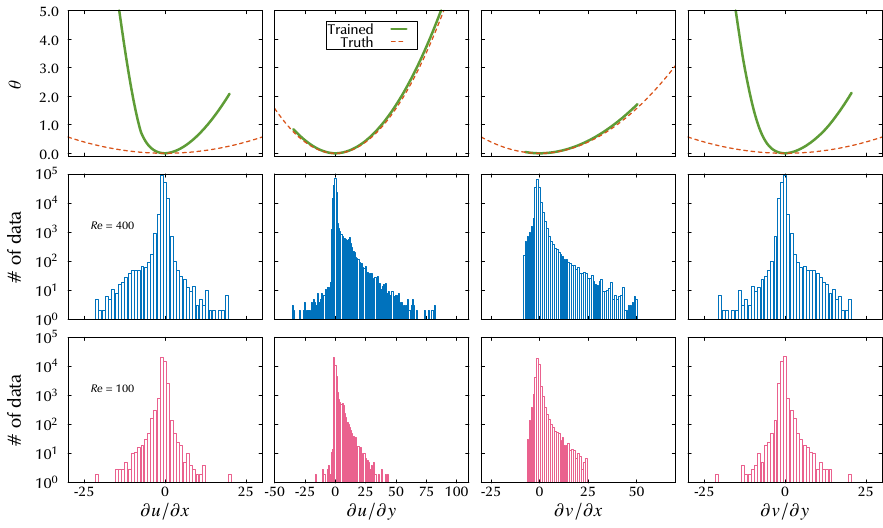} \\
    (a) & (b)
    \end{tabular}   
    \caption{(a) Comparison between true and trained dissipation functions plotted for the dataset A.
    (b) Dissipation function projected to one dimension (upper) and data distributions corresponding to the projected axes (middle and lower).}
    \label{fig-nonNewton2}
\end{figure}

\subsubsection*{Results}
Results for model discovery and generalization check are presented in \cref{fig-nonNewton1,fig-nonNewton2}.
The upper panels in \cref{fig-nonNewton1} show the results of the model discovery stage, where DissipationNet is trained.
It can be clearly seen that the loss function (\cref{fig-nonNewton1}a) converged down to the order of \num{1e-8}, and the predicted flow field (\cref{fig-nonNewton1}b) showed good agreement (\cref{fig-nonNewton1}d) with the reference solution (\cref{fig-nonNewton1}c).
The maximum absolute errors between the predicted and reference solutions are \num{5e-3} for model discovery (\cref{fig-nonNewton1}d) and \num{1e-2} for model application (\cref{fig-nonNewton1}h).

The discovered dissipation function $\theta$ is shown in \cref{fig-nonNewton2} with histograms of $\nabla\bol{u}$ involved in the datasets.
\Cref{fig-nonNewton2}a shows a comparison between true and trained dissipation functions plotted for the dataset A, where the correlation coefficient is evaluated as \num{0.99975}.
It can be seen that $\theta$ is well-defined for the range of velocity gradients $\nabla\bol{u}$ in the dataset A.
For the range outside of the dataset, the dissipation function $\theta$ still keeps good accuracy for $u_y$ and $v_x$. 
However, for $u_x$ and $v_y$, the applicable range is quite narrow.
The dataset A is the lid-driven cavity flow, where the moving top wall generates strong shear.
This shear appears in $u_y$ and $v_x$ in the coordinate system employed.

For the generalization check, the trained DissipationNet is fixed and is applied to the dataset B, which has a different Reynolds number from the dataset A.
It is confirmed that the loss function converged sufficiently (\cref{fig-nonNewton1}e), the predicted solution (\cref{fig-nonNewton1}f) showed good agreement (\cref{fig-nonNewton1}h) with the reference solution calculated by \texttt{OpenFOAM} (\cref{fig-nonNewton1}g).
Regarding the difference of the Reynolds number, the coefficient $\Gamma_\Rey$ converged close to the true value $\Rey_A / \Rey_B = 4$ with sufficiently small relative error \SI{0.07}{\percent}.

\section{Summary and Discussion}\label{sec-conclusion}
We introduced \emph{GIMLET}, a novel framework for gray-box model discovery that is generalizable, interpretable, and library-free. By simultaneously satisfying these three criteria, the proposed methodology represents a significant advancement in data-driven and physics-informed model discovery. The validity of \emph{GIMLET} was demonstrated across several benchmark problems, and its strong generalizability was verified for a diverse range of systems.

To clarify the advantage of the present \emph{GIMLET}, the model discovery is compared with PINN-SR \cite{Chen2021}, which is a state-of-the-art symbolic regression type of model discovery.
As illustrated in \ref{app-PINN_SR}, \emph{GIMLET} can appropriately discover the constitutive laws even from noisy data for which the PINN-SR fails to identify the correct term.

The central idea of the proposed approach is to recover unknown terms in flow transport equations through functional derivatives of the free-energy and/or dissipation functions. This formulation enables the extraction of highly generalizable models directly from data, while reducing reliance on local or data-specific solution behaviors. Furthermore, interpretability is inherently ensured by the thermodynamic structure of the method: the inferred free-energy and dissipation functions provide direct physical insight into the learned dynamics. Finally, the library-free nature of \emph{GIMLET} offers a substantial practical advantage over existing symbolic regression-based approaches, particularly for real-world applications in data-driven physics model discovery.

One may argue that the trained thermodynamic functions expressed by neural networks are still `gray-box'.
After the thermodynamic functions are trained by \emph{GIMLET}, the mathematical expression can be obtained by applying symbolic regressions (SRs) based on evolutionary algorithms and, as in SINDy, based on the assumption of model sparsity. Applying SR to the learned potentials is simpler than applying it directly to the spatial-derivative terms of the governing equations, because $g$ and $\theta$ are low-dimensional scalar functions of local state variables.

In this study, the scope of model discovery is restricted to constitutive relations embedded within partial differential equations governing fluid flow and scalar transport. 
This restriction, however, is not fundamental. 
The \emph{GIMLET} framework can be naturally extended to a broader class of physical phenomena by introducing additional field variables that contribute to the system's free-energy functional. 
Potential extensions include constitutive modeling of viscoelastic materials, electro- and magnetohydrodynamic systems, and multiphase flows. 
Extending the framework to transport phenomena involving phase changes may require addressing the challenge of discontinuities in the free-energy landscape. 
Furthermore, the present formulation assumes simple boundary conditions within the variational principle. 
By relaxing this assumption, the proposed approach could be extended to the data-driven discovery of boundary conditions themselves.

In \emph{GIMLET}, the free energy and dissipation functions are represented by neural networks.
In the benchmark problems treated in this study, the functions to be discovered are fairly simple, except for the non-Newtonian fluid case. 
For the Bird--Cross--Carreau--Yasuda model \cref{eq-BCCY}, the effective viscosity is highly nonlinear, which is hard to precisely fit by simple functions, e.g., a polynomial.
In addition, as mentioned above, the \emph{GIMLET} can be extended to more complicated phenomena, for instance, viscoelastic fluids. 
To this end, the representation of the free energy and dissipation functions by neural networks that have sufficient capability is a reasonable design.

Finally, although the computational experiments in this work are limited to two-dimensional settings---consistent with the two-dimensional nature of standard PIV measurements---the proposed methodology is not inherently dimension-dependent. It can be extended to three-dimensional problems and, in principle, enables three-dimensional field reconstruction from two-dimensional observational data. We expect that the present framework will also pave the way for data-driven discovery of generalizable closures for complex turbulent flows.

\section*{Declaration of competing interest}
The authors declare that they have no known competing financial interests or personal relationships that may have influenced the work reported in this study.

\section*{Acknowledgements}
The work of S.S. has been supported by the grant-in-aid KAKENHI JP23KK0262.
The work of E.K., K.S., and G.E.K. has been supported by the DOE-MMICS SEA-CROGS DE-SC0023191 and, in part by AIM for Composites, an Energy Frontier Research Center funded by the U.S. Department of Energy (DOE), Office of Science, Basic Energy Sciences (BES), under Award No. DE-SC0023389.
The authors are grateful to Dr. Zhicheng Wang of Brown University for providing data for the flow past a cylinder.
S.S. expresses appreciation for the support from M.D. Yuu Kawashima.

\section*{Data availability}
Code and datasets used in this study are available on GitHub at the following: \\
\url{https://github.com/sugurushiratori/GIMLET} (will be available after publication).

\appendix

\section{Hyperparameters}\label{app-params}
The hyperparameters used in the validations are summarized in \cref{tab-hyper_parms}.
The network sizes are determined based on the survey described in the next section.

\begin{table}[tp]
\caption{\label{tab-hyper_parms} The hyperparameters used in the validation cases.}
\small
\rowcolors{2}{tb1}{white}
\begin{tabularx}{\textwidth}{lccccccccC}
\toprule
 \rowcolor{tbH} \head{Equations} & \head{Sec.} 
 & \head{$\beta_{\ell_2}$} & \head{$\beta_g$} & \head{$\beta_\theta$} & \head{$\beta_h$}
 & \head{$N_f$} & \head{$N_d$} 
 & \head{Adam} & \head{SSBroyden}
 \\
\midrule
Burgers                       & \ref{sec-Burgers}  & \num{1e-11} & 0.0 & 0.0 & 0.0 & \num{5000} & \num{5000} & \num{1000} & \num{10000} \\
Kuramoto--Sivashinsky          & \ref{sec-KS}       & \num{1e-11} & 1.0 & 0.0 & 0.0 & \num{30000} & \num{5000}  & \num{1000} & \num{200000} \\
Navier--Stokes (Newtonian)     & \ref{sec-NS}       & \num{1e-11} & 0.0 & 0.0 & 1.0 & \num{20000} & \num{10000}  & \num{1000} & \num{100000}  \\
Navier--Stokes (non-Newtonian) & \ref{sec-NonNewton}& \num{1e-11} & 0.0 & 0.0 & 1.0 & \num{30000} & \num{30000}  & \num{1000} & \num{100000}  \\
\bottomrule
\end{tabularx}
\end{table}

\subsection*{Effect of network sizes}
Next, the dependence of training results on the size of neural networks is investigated for the case of the Burgers equation.
In the present study, the neural network is composed of fully connected hidden layers; the network size is simply determined by the number of hidden layers $N_\ell$ and the number of neurons $N_h$ in each layer.
For the Burgers equation, the \emph{GIMLET} architecture has two neural networks: a PINN and a DissipationNet. 
\Cref{tab-SizeDepP1} shows final values of the loss function for different combinations of $N_\ell$ and $N_h$ of the PINN network.
The size of the DissipationNet is kept fixed as $N_\ell\times N_h = 4\times 10$.
For the larger values of $N_\ell$ and $N_h$, the final loss functions $\mathcal{J}$ decrease, 
and $\mathcal{J}$ becomes less dependent for sufficiently large networks.
The combination $N_\ell\times N_h = 8\times 20$, which is employed in \cref{sec-Burgers}, can be regarded as sufficiently large.
To evaluate the accuracy of model discovery, the root mean square relative error (RMSRE) in the dissipation function $\theta(u_x)$ is evaluated for different PINN sizes, and the results are summarized in \cref{tab-SizeDepP2}.
It can be seen that the dissipation function is accurately discovered even for insufficient PINN sizes.

\begin{table}[tbp]
\caption{\label{tab-SizeDepP1} Final loss function for different network sizes of the PINN.}
\small
\centering
\begin{tabularx}{0.6\textwidth}{lcCCC}
\toprule
  & &  \multicolumn{3}{c}{$N_h$} \\
  & &   \num{10} & \num{20} & \num{30} \\
\midrule
\multirow{3}{*}{$N_\ell$}
 &  4 & \num{1.095e-07} & \num{7.036e-11} & \num{3.803e-11} \\
 &  6 & \num{5.850e-10} & \num{3.103e-11} & \num{2.897e-11} \\
 &  8 & \num{1.913e-10} & \num{2.743e-11} & \num{2.912e-11} \\
\bottomrule
\end{tabularx} 
\end{table}

\begin{table}[tbp]
\caption{\label{tab-SizeDepP2} Root mean squared relative errors (RMSRE) in dissipation function $\theta$ for different network sizes of the PINN.}
\small
\centering
\begin{tabularx}{0.6\textwidth}{lcCCC}
\toprule
  & &  \multicolumn{3}{c}{$N_h$} \\
  & &  \num{10} & \num{20} & \num{30} \\
\midrule
\multirow{3}{*}{$N_\ell$}
 &  4 & \SI{0.35617}{\percent} & \SI{0.35673}{\percent} & \SI{0.35674}{\percent} \\
 &  6 & \SI{0.35667}{\percent} & \SI{0.35675}{\percent} & \SI{0.35673}{\percent} \\
 &  8 & \SI{0.35673}{\percent} & \SI{0.35674}{\percent} & \SI{0.35674}{\percent} \\
\bottomrule
\end{tabularx}
\end{table}

Next, the effect of the DissipationNet size is investigated with fixed PINN size as $N_\ell\times N_h = 8\times 20$.
\Cref{tab-SizeDepG1} shows the final values of the loss function for different combinations of $N_\ell$ and $N_h$ of the DissipationNet.
In contrast to the PINN sizes (\cref{tab-SizeDepP1}), the loss functions are almost the same in the shown range, even for small networks  $N_\ell\times N_h = 2\times 5$.
The accuracy of the model discovery, which is shown in \cref{tab-SizeDepG2}, is also almost constant for different sizes of DissipationNet.

\begin{table}[tbp]
\caption{\label{tab-SizeDepG1} Final loss function for different network sizes.}
\small
\centering
\begin{tabularx}{0.6\textwidth}{lcCCC}
\toprule
  & &  \multicolumn{3}{c}{$N_h$} \\
  & &   \num{5} & \num{10} & \num{15} \\
\midrule
\multirow{3}{*}{$N_\ell$}
 &  2 & \num{3.056e-11} & \num{3.022e-11} & \num{3.337e-11} \\
 &  4 & \num{2.884e-11} & \num{2.743e-11} & \num{2.764e-11} \\
 &  6 & \num{2.755e-11} & \num{2.866e-11} & \num{2.734e-11} \\
\bottomrule
\end{tabularx}
\end{table}

\begin{table}[tbp]
\caption{\label{tab-SizeDepG2} Root mean squared relative errors (RMSRE) in dissipation function $\theta$ for different network sizes.}
\small
\centering
\begin{tabularx}{0.6\textwidth}{lcCCC}
\toprule
  & &  \multicolumn{3}{c}{$N_h$} \\
  & &   \num{5} & \num{10} & \num{15} \\
\midrule
\multirow{3}{*}{$N_\ell$}
 &  2 & \SI{0.35675}{\percent} & \SI{0.35674}{\percent} & \SI{0.35674}{\percent} \\
 &  4 & \SI{0.35674}{\percent} & \SI{0.35674}{\percent} & \SI{0.35672}{\percent} \\
 &  6 & \SI{0.35674}{\percent} & \SI{0.35674}{\percent} & \SI{0.35673}{\percent} \\
\bottomrule
\end{tabularx}
\end{table}

\subsection*{Effect of random initialization of neural networks}
For general neural networks, training results may vary with initial weights and biases, which may lead to failure of the model discovery.
To check the robustness of the \emph{GIMLET}, the effect of random seeds on the training results is investigated for the case of Burgers equations, which is shown in \cref{sec-Burgers}.
10 runs were performed with different random seeds, and the values of the total loss function were \num{2.87e-11} on average with sufficiently small standard deviation \num{1.68e-12}.
For the discovered dissipation function $\theta$, the root mean squared relative errors (RMSRE) between the discovered and exact functions were \SI{0.3567}{\percent} with a standard deviation of \SI{2.804e-8}{\percent}, which indicates that the dependence of discovered models on the random seed is sufficiently small.

\section{Comparison with PINN-SR for noisy data}\label{app-PINN_SR}
To clarify the advantage of the present \emph{GIMLET}, the model discovery is compared with PINN-SR \cite{Chen2021}, which is a state-of-the-art symbolic regression type of model discovery.
The comparison is also carried out for noisy data to check the practical capability of the \emph{GIMLET}.
While the original PINN-SR is designed to identify all terms in the PDE except the time-derivative term, the \emph{GIMLET} is designed to discover the constitutive laws only.
To deal with this, we modified it so that only the missing term can be identified.
In the following, the method of modified PINN-SR is briefly explained, and then the numerical results for the Burgers equation are shown.

\subsection*{Modified PINN-SR for missing term discovery}
In the modified PINN-SR, the solution $u$ is predicted by a PINN, which is parametrized by $\bol{\lambda}$, and the governing equation is expressed as 
\begin{equation}
\mathcal{F}_\text{known}
 \Big[ u\paren{\bol{\lambda}} \Big]  + \bol{\Phi}\paren{\bol{\lambda}} \cdot \bol{\Lambda} = 0,
\end{equation}
where $\mathcal{F}$ represents known terms in the governing equation, and the missing term is represented by the inner product of two vectors $\bol{\Phi}$ and $\bol{\Lambda}$. 
$\bol{\Phi}$ is a so-called function library, which is composed of candidate functions of spatial derivatives. $\bol{\Lambda}$ is the unknown coefficients.
For the Burgers equation treated below, the known terms are $\mathcal{F}_\text{known} = u_t + u u_x$, and the library is composed of the following 16 candidates:
\begin{equation}
\bol{\Phi} = \left[\,
  1,\; u,\; u^2,\; u^3,\;
  u_x,\; u u_x,\; u^2 u_x,\; u^3 u_x,\;
  u_{xx},\; u u_{xx},\; u^2 u_{xx},\; u^3 u_{xx},\;
  u_{xxx},\; u u_{xxx},\; u^2 u_{xxx},\; u^3 u_{xxx}
\,\right]^t .
\end{equation}
Since the term to be discovered is the diffusion term $\eta u_{xx}$, a successful identification corresponds to a single nonzero coefficient for $u_{xx}$, and the number of misspecified terms reported in \cref{tab-PINN_SR} counts the remaining nonzero components of $\bol{\Lambda}$.
To find sparse regression of the coefficients $\bol{\Lambda}$, we consider a minimization of the total loss function defined as
\begin{equation}
  \left(\bol{\lambda}^*, \bol{\Lambda}^* \right) =
 \underset{\bol{\lambda}, \bol{\Lambda}}{\text{argmin}}
 \left[ \mathcal{J}_d\paren{\bol{\lambda}; \mathcal{D}_m}
  + \alpha \mathcal{J}_p^\text{SR}\paren{\bol{\lambda}, \bol{\Lambda}; \mathcal{D}_f}
  + \beta_{\ell_0} \left\| \bol{\Lambda} \right\|_0
  \right],  \label{eq-PINN_SR_loss}
\end{equation}
where
$\mathcal{J}_d$ and $\mathcal{J}_p^\text{SR}$ are the data loss and physics loss, respectively.
$\alpha$ and $\beta_{\ell_0}$ denote the relative weighting of the loss functions, while $\mathcal{D}_m$ and $\mathcal{D}_f$ represent the observation data and residual points, respectively.
The data loss is defined as the mean squared error between predicted and observed solution, the same as \cref{eq-JdDef}.
The physics loss is defined as the mean-squared error in the governing equation:
\begin{equation}
\mathcal{J}_p^\text{SR}\paren{\bol{\lambda}, \bol{\Lambda}; \mathcal{D}_f}
 = \frac{1}{N_f} \bigg\| \mathcal{F}_\text{known}
 \Big[ u\paren{\bol{\lambda}} \Big]  + \bol{\Phi}\paren{\bol{\lambda}} \cdot \bol{\Lambda}\bigg\|^2_2.
\end{equation}
A minimization \cref{eq-PINN_SR_loss} is highly intractable since the $\ell_0$ regularization makes this problem NP-hard. 
According to the original PINN-SR \cite{Chen2021}, the alternating direction optimization (ADO) is applied. The ADO divides the overall optimization problem into a set of tractable problems:
\begin{subequations}
\begin{align}
\bol{\Lambda}^*_{k+1} &= \underset{\bol{\Lambda}}{\text{argmin}}
 \left[ \mathcal{F}_\text{known}\Big[ u \paren{\bol{\lambda}^*_k} \Big] + \bol{\Phi}\paren{\bol{\lambda}^*_k}\cdot\bol{\Lambda}
  + \beta_{\ell_0} \left\| \bol{\Lambda} \right\|_0
  \right] \label{eq-ADO1} \\
\bol{\lambda}^*_{k+1} &= \underset{\bol{\lambda}}{\text{argmin}}
 \left[ \mathcal{J}_d\paren{\bol{\lambda}; \mathcal{D}_m}
  + \alpha \mathcal{J}_p^\text{SR}\paren{\bol{\lambda}, \bol{\Lambda}^*_{k+1}; \mathcal{D}_f} \right]
  \label{eq-ADO2}
\end{align}
\end{subequations}
In each ADO iteration, the coefficients $\bol{\Lambda}$ are updated via STRidge (a sequential thresholding regression~\cite{Brunton2016,Rudy2017}).
The neural network parameters $\bol{\lambda}$ are then updated by a standard backpropagation algorithm.
Before running the ADO iteration, the pre-training is conducted by replacing the regularization term from $\ell_0$ to $\ell_1$, namely,
\begin{equation}
  \left(\bol{\lambda}^*, \bol{\Lambda}^* \right) =
 \underset{\bol{\lambda}, \bol{\Lambda}}{\text{argmin}}
 \left[ \mathcal{J}_d\paren{\bol{\lambda}; \mathcal{D}_m}
  + \alpha \mathcal{J}_p^\text{SR}\paren{\bol{\lambda}, \bol{\Lambda}; \mathcal{D}_f}
  + \beta_{\ell_1} \left\| \bol{\Lambda} \right\|_1
  \right],
\end{equation}
where $\beta_{\ell_1}$ denotes the weight for $\ell_1$ regularization.
After the ADO iteration, a post-training is done by the following minimization, where the regularization is removed.
\begin{equation}
  \left(\bol{\lambda}^*, \bol{\Lambda}^* \right) =
 \underset{\bol{\lambda}, \bol{\Lambda}}{\text{argmin}}
 \left[ \mathcal{J}_d\paren{\bol{\lambda}; \mathcal{D}_m}
  + \alpha \mathcal{J}_p^\text{SR}\paren{\bol{\lambda}, \bol{\Lambda}; \mathcal{D}_f} \right].
\end{equation}

\subsection*{Results for Burgers equation}
Both the \emph{GIMLET} and the modified PINN-SR are conducted for the case of the Burgers equation, which has been shown in \cref{sec-Burgers}.
The size of the neural networks is set at $N_\ell\times N_h = 8\times 20$ for both methods.
Regarding the optimizer, the SSBroyden is also used for the PINN-SR.
In the PINN-SR, it is known that the value of $\alpha$, which is the weighting factor for the physics loss, sensitively affects the results.
Here, the $\alpha$ is selected as $\alpha=0.7$ for dataset A and $\alpha=0.2$ for dataset B. 
In the pretraining stages, the value of $\alpha$ is set to half of that used in the main ADO loop.

Results of model discovery are summarized in \cref{tab-PINN_SR}.
Model discoveries are carried out for two datasets, which are treated in \cref{sec-Burgers}. Dataset A involves a steeper spatial derivative than dataset B. 

As can be seen from \cref{tab-PINN_SR}, the PINN-SR failed for most cases, except for the noise level of \SI{0}{\percent} in Dataset B.
Although in the original PINN-SR paper~\cite{Chen2021}, the Burgers equation is also treated and succeeded in discovering all the terms, note that the viscosity of the present case ($\eta=0.01 / \pi$) is much smaller than that used in \cite{Chen2021} ($\eta=0.1$).
Among the failure cases of PINN-SR, errors ($\epsilon_\text{pred}$ and $\epsilon_\text{mis}$) are very small for the case of noise \SI{0}{\percent} of dataset A. In this case, the coefficient for the correct viscous term $u_{xx}$ is accurate, and the six coefficients for the misspecified terms are very small.
Thus, the discovered model is expected to be applied to different configurations of the same governing equation.
For other failure cases of the modified PINN-SR, the coefficients for the misspecified terms are quite large, which means they are not generalizable to other data.

\begin{table}[tbp]
\caption{\label{tab-PINN_SR} Summary of model discovery results in comparison with PINN-SR \cite{Chen2021} for the case of Burgers equation.
For the \emph{GIMLET}, the accuracy is evaluated by root mean squared relative errors (RMSRE) between discovered and exact dissipation function $\theta$.
For the PINN-SR, the number of misspecified terms (\# mis-spec.) is shown for failure cases.
The accuracy is evaluated as relative error $\epsilon_\text{pred}$ in the coefficient of the diffusion term $u_{xx}$.
For failure cases, the root mean squares of normalized coefficients $\epsilon_\text{mis}$ are also evaluated.
The \emph{GIMLET} was also run for the noise level of \SI{1}{\percent}, and the discovered dissipation function is included in \cref{fig-noise_Burgers_uG}; the PINN-SR was not run for that noise level, and it is therefore not listed in this comparison.}
\small
\centering
\begin{tabularx}{\textwidth}{ccCcCCC}
\toprule
 & &  GIMLET  && \multicolumn{3}{c}{PINN-SR} \\
    \cline{5-7} 
 & &  RMSRE   &&  S/F (\# mis-spec.) & $\epsilon_\text{pred}$ & $\epsilon_\text{mis}$  \\

\midrule
\multirow{3}{*}{Dataset A}
& noise \SI{0}{\percent}  & \SI{0.356}{\percent} 
&& Fail (6)  & \SI{0.01}{\percent} & \SI{0.013}{\percent} \\
& noise \SI{5}{\percent}  & \SI{0.366}{\percent} 
&& Fail (13) & \SI{19.9}{\percent} & \SI{348}{\percent}\\
& noise \SI{10}{\percent} & \SI{0.368}{\percent} 
&& Fail (14) & \SI{29.6}{\percent} & \SI{809}{\percent} \\
\midrule
\multirow{3}{*}{Dataset B}
& noise \SI{0}{\percent}  &\SI{0.420}{\percent} 
&& Success & \SI{0.001}{\percent} & --- \\
& noise \SI{5}{\percent}  &\SI{0.432}{\percent} 
&& Fail (2) & \SI{0.447}{\percent}& \SI{7.42}{\percent}\\
& noise \SI{10}{\percent} &\SI{0.672}{\percent} 
&& Fail (10)& \SI{8.17}{\percent} & \SI{1055}{\percent}\\
\bottomrule
\end{tabularx}
\end{table}

\begin{figure}[tbp]
    \centering
    \begin{tabular}{cc}
    \includegraphics[width=0.48\textwidth]{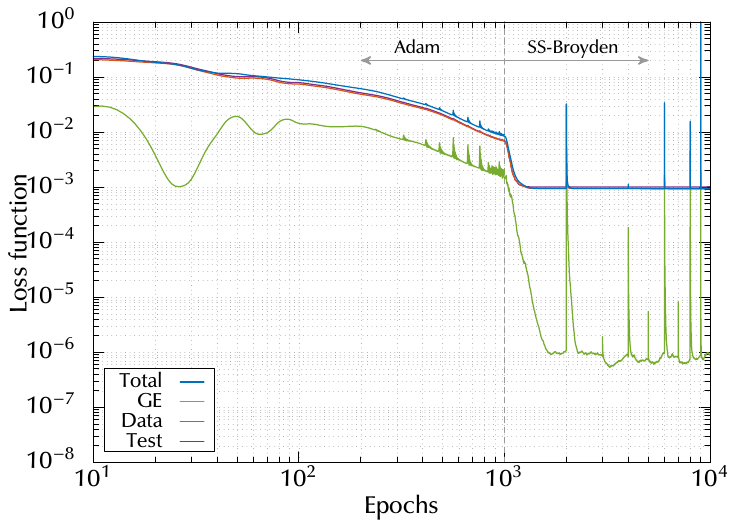} & 
    \includegraphics[width=0.48\textwidth]{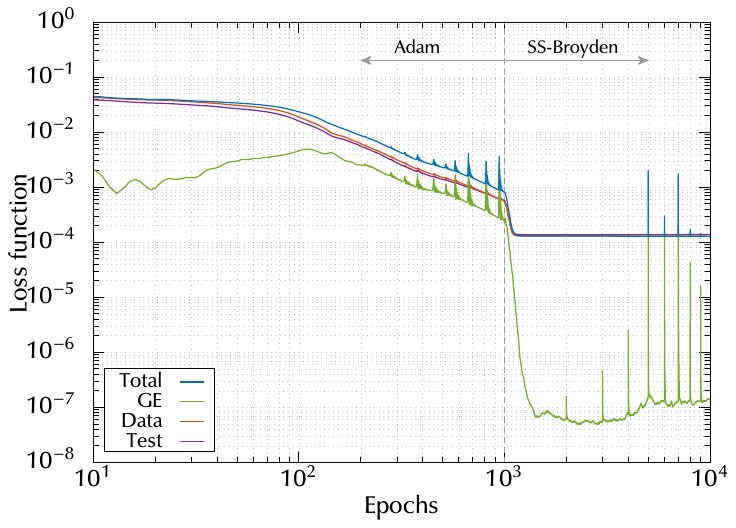} \\
    (a) Dataset A & (b) Dataset B
    \end{tabular}   
    \caption{History of loss functions for the discovery of diffusion term in Burgers equation with noise of \SI{5}{\percent}. Results for Dataset A (a)  and Dataset B (b).
    \label{fig-noise_Burgers_loss}}
\end{figure}

\begin{figure}[tbp]
    \centering
    \begin{tabular}{cc}
    \includegraphics[width=0.48\textwidth]{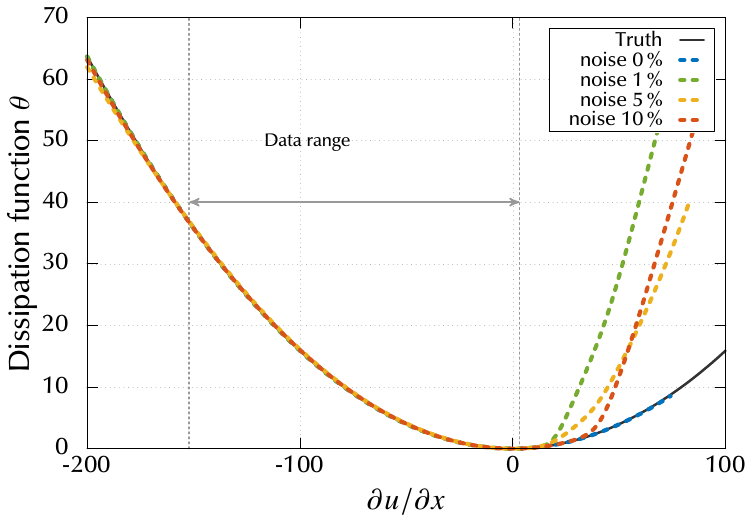} & 
    \includegraphics[width=0.48\textwidth]{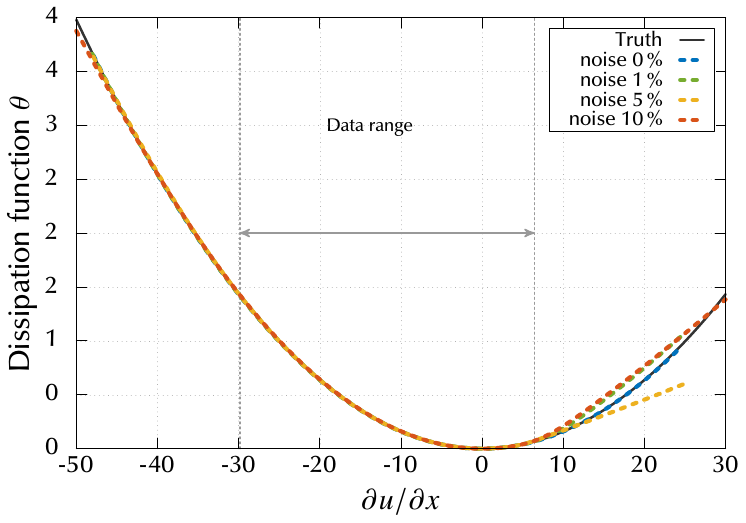} \\
    (a) Dataset A & (b) Dataset B
    \end{tabular}   
    \caption{Dissipation functions $\theta\paren{u_x}$ discovered by the \emph{GIMLET} for different noise levels. Results for Dataset A (a)  and Dataset B (b).
    Gray solid lines are ground truth, whereas dashed color lines are $\theta\paren{u_x}$ discovered by the \emph{GIMLET}.
    \label{fig-noise_Burgers_uG}}
\end{figure}

On the other hand, the \emph{GIMLET} succeeded in discovering the dissipation function even for noisy data, though the RMSRE gets worse as noise increases.
To see further how the \emph{GIMLET} works for the noisy data, let us check the loss functions and discovered dissipation functions.
\Cref{fig-noise_Burgers_loss}(a,b) shows the history of loss functions during training of the \emph{GIMLET} for datasets A and B, respectively.
For both cases, the data loss stopped reducing at the same order of magnitude of the given noise level, whereas the physics loss can be further reduced from 3 to 5 orders of magnitude.
The discovered dissipation functions $\theta(u_x)$ are plotted in \cref{fig-noise_Burgers_uG}(a,b) for both cases of dataset A and B, respectively, for noise levels of 0, 1, 5 and \SI{10}{\percent}.
It can be seen that the discovered $\theta(u_x)$ are almost perfectly in agreement with the truth, even for noisy data, at least in the range of $u_x$ that is involved in the data.
Even for the outside of the data range, the discovered function $\theta$ is still correct, especially in the negative $u_x$ range.
The discovered $\theta(u_x)$ becomes less accurate for a positive $u_x$ range, especially for dataset A, where the data has a steep spatial gradient.
This inaccuracy might be improved by applying data augmentation, which is described in \cref{sec-argument}.

\pagebreak

\section{Variational principle}\label{app-VP}
The derivation of basic forms of governing equations is described in this section.
We modify the variational principle proposed by Fukagawa and Fujitani \cite{Fukagawa2010,Fukagawa2012}, keeping free energy and dissipation function unknown.

\subsection*{Lagrangian and Eulerian descriptions}
We consider the dynamics of this fluid in a fixed container 
from the initial time $t_\text{ini}$ to the final time $t_\text{fin}$.
Let $V$ and $\partial V$ denote the region occupied by the fluid and its boundary, respectively.

In the Lagrangian description, we label a fluid particle with its initial position
$\bol{a} = (a_1, a_2, a_3)$, and write $\bol{X} = (X_1, X_2, X_3)$ for its position at time $\tau$.
The time derivative of $\bol{X}$ denotes the velocity fields $\bol{u}$, i.e.,
\begin{equation}
  \partial_\tau\bol{X} = \bol{u}.
\end{equation}
The endpoints of a path line are fixed by
\begin{equation}
   \delta \bol{X}(\bol{a}, t_\text{ini} ) =
   \delta \bol{X}(\bol{a}, t_\text{fin} ) = \bol{0}.
\end{equation}
The fluid particles may expand or shrink along the flow path.
The expansion rate, which is also called the volume element, can be given by the determinant of the Jacobian matrix
\begin{equation}
  J \left( \bol{a}, \tau \right) = \pdif{(X_1, X_2, X_3)}{(a_1, a_2, a_3)}.
\end{equation}
We assume that $J$ has no singular points in the space and time considered.
From $\bol{a} = \bol{X}(t_\text{ini}, \bol{a})$, $J(\bol{a}, t_\text{ini}) = 1$ is given.

In the Eulerian description, a fluid particle is labeled by the Lagrangian coordinates
$\bol{A} = (A_1, A_2, A_3)$, which depend on the spatial position $\bol{x}$.
Here $\bol{A}$ is a function that gives the Lagrangian coordinates from the Eulerian coordinates.
Since the Lagrangian coordinates are conserved along the path line, we have
\begin{equation}
   \pdif{A_i}{t} + \bol{u}\cdot\nabla A_i = 0,  \label{App_eq-A}
\end{equation}
and the endpoints of a path line are fixed by
\begin{equation}
   \delta \bol{A}(\bol{x}, t_\text{ini} ) =
   \delta \bol{A}(\bol{x}, t_\text{fin} ) = \bol{0}.   \label{App_eq-constA}
\end{equation}
The mass conservation law can be written as
\begin{equation}
   \rho(\bol{x},t) - J^{-1}(\bol{x},t) \rho_\text{ini} \left(\bol{A}(\bol{x},t) \right)  = 0,    \label{App_eq-constRho}
\end{equation}
where $J^{-1}$ is given by
\begin{equation}
  J^{-1} \left( \bol{x}, t \right) = \pdif{(A_1, A_2, A_3)}{(x_1, x_2, x_3)}.
\end{equation}

\subsection*{Formulation}
We consider the dynamics of a viscous two-component fluid.
Let $\rho_a$ and $\rho_b$ be the mass densities of components $a$ and $b$, respectively.
The total mass density and the mass fraction of component $a$ are given by
\begin{equation}
   \rho \equiv \rho_a + \rho_b, \qquad \phi \equiv \frac{\rho_a}{\rho}.
\end{equation}
The chemical potential is defined as the Gibbs energy per unit mass,
and $\mu\paren{\phi}$ is the difference of chemical potential between two components.
In addition to the free energy due to the fluid motion, the free energy due to the two-component mixture is defined as $g(\phi)$, and we want to discover the form of $g\paren{\phi}$ from the observed data.
The total additional free energy of this system is defined as
\begin{equation}
   G[\phi] \equiv \int_V g(\phi) dV,
\end{equation}
and the chemical potential $\mu$ can be written as
\begin{equation}
   \mu \equiv \frac{1}{\rho}\frac{\delta G}{\delta \phi},
\end{equation}
where $\delta / \delta \phi$ is the functional derivative.
The specific internal energy $\epsilon$ is a function of the density $\rho$, the entropy $s$, and the mass fraction $\phi$ as
\begin{equation}
   p   \equiv \rho^2 \left( \pdif{\epsilon}{\rho} \right)_{s,\phi}, \qquad
   T   \equiv \left( \pdif{\epsilon}{s}   \right)_{\rho,\phi},      \qquad
   \mu \equiv \left( \pdif{\epsilon}{\phi} \right)_{s,\rho},
\end{equation}
where the subscripts indicate variables fixed in the respective partial differentiations.
From thermodynamics, the total differentiation of $\epsilon$ can be written as
\begin{equation}
   d\epsilon =  \frac{p}{\rho^2}d\rho + Tds + \mu d\phi.
\end{equation}

The mass conservation of the components $a$ and $b$ can be given as
\begin{equation}
  \partial_t \rho_a + \nabla\cdot\left( \rho_a \bol{u} + \bol{j}_a  \right) = 0, \qquad
  \partial_t \rho_b + \nabla\cdot\left( \rho_b \bol{u} + \bol{j}_b  \right) = 0,  \label{App_eq-mass_ab}
\end{equation}
where $\bol{j}_a$ and $\bol{j}_b$ are diffusive fluxes, which satisfy $\bol{j}_a = -\bol{j}_b$.
\Cref{App_eq-mass_ab} can be rewritten as
\begin{gather}
  \partial_t \rho + \nabla\cdot\left( \rho \bol{u}\right) = 0, \\
  \rho\left( \partial_t \phi + \left(\bol{u}\cdot\nabla\right)\phi \right) = - \nabla\cdot\bol{j}_a, \label{App_eq-phiTrans}
\end{gather}

We denote $\check{\sigma}_{ij}$ and $\bol{q}$ for the viscous stress tensor and the heat flux, respectively.
And, we define the rate-of-strain tensor $e_{ij}$ as
\begin{equation}
  e_{ij} = \frac{1}{2} \left( \pdif{u_i}{x_j} + \pdif{u_j}{x_i} \right).
\end{equation}

For dissipative processes, the time evolution of entropy can be written as
\begin{equation}
   \rho T \left( \pdif{s}{t} + \bol{u}\cdot\nabla s \right)  - \theta
   + \nabla\cdot \left( \bol{q} + \mu\bol{j}_a \right)= 0, 
   \label{App_eq-constraintNH}
\end{equation}
where $\theta$ is the dissipation function, which expresses how the frictional force converts the energy into heat.
The dissipation function $\theta$ can be written as
\begin{equation}
  \theta = \check{\sigma}_{ij} e_{ij} + \nu D_t \phi,
\end{equation}
where the first term on the right-hand side is dissipation due to viscous motion.
The second term expresses that the dissipation is caused by the change of mass fraction $\phi$ along the fluid particle path, and $\nu$ is the friction coefficient.
The form of the dissipation function $\theta(\bol{u}, \phi)$ is also a target for data-driven discovery.
The total dissipation potential is defined as
\begin{equation}
   \Theta[\bol{u},\phi] \equiv \int_V \theta(\nabla\bol{u},\nabla\phi) dV.
\end{equation}

\subsection*{Principle of least action}
Next, we derive the governing equations by considering the principle of least action.
We define the Lagrangian density as
\begin{equation}
   \mathcal{L}(\rho, \bol{u}, s, \phi)  \equiv \rho \left\{ \frac{1}{2}\bol{u}^2  - \epsilon(\rho,s,\phi) \right\}
\end{equation}
Holonomic constraints \cref{App_eq-constA,App_eq-constRho} can be considered in the principle of least action path, using Lagrange multipliers as 
\begin{equation}
I = \int_{t_\text{ini}}^{t_\text{fin}} dt \int_V dV
\{ \mathcal{L} + \beta_i \left( \partial_t A_i + \bol{u}\cdot\nabla A_i \right) + K\left( \rho - \rho_\text{ini}J^{-1} \right)   \},   \label{App_eq-action}
\end{equation}
which is a functional of $\bol{\beta}$, $K$, $\rho$, $s$, $\bol{u}$, $\bol{A}$ and $\phi$.
The time evolution of entropy \cref{App_eq-constraintNH}, which is a non-holonomic constraint, must be considered separately, because the Lagrange multiplier method cannot be applied.
We assume the no-slip condition for velocity field $\bol{u}$ at the boundary $\partial V$, i.e.,
\begin{equation}
  \bol{u} = 0  \qquad \text{on} \qquad \partial V,
\end{equation}
and thus, we can assume no dissipation at the boundary.
In addition, we assume that the whole fluid is enclosed by an adiabatic wall, i.e., that no heat flux can pass through the boundary.
By integrating \cref{App_eq-constraintNH} over $V$, we obtain
\begin{equation}
  \begin{split}
  & \int_V dV \left\{ \rho T \left( \partial_t s + \bol{u}\cdot\nabla \right) s - \theta
   + \nabla\cdot \left( \bol{q} + \mu\bol{j}_a \right)  \right\} \\
= & \int_V dV \left\{ \rho T \left( \partial_t s + \bol{u}\cdot\nabla \right) s
 - \check{\sigma}_{ij} \pdif{u_j}{x_i} - \nu\left( \partial_t\phi + \bol{u}\cdot\nabla \right) \phi \right\} \\
= & \int_V dV \left\{ \rho T \partial_t s - \nu\partial_t\phi-
  \pdif{x_j}{A_i} \left( \rho T \pdif{s}{x_j} - \nu\pdif{\phi}{x_j} + \check{f}_j \right) \pdif{A_i}{t} \right\} = 0
   \end{split}
\end{equation}
where the relations and definition
\begin{equation}
  \check{\sigma}_{ij} e_{ij} = \check{\sigma}_{ij} \pdif{u_j}{x_i},  \qquad
   u_j = - \pdif{x_j}{A_i}\pdif{A_i}{t}, \qquad
   \int_V dV \left( \check{\sigma}_{ij} \pdif{u_j}{x_i} \right)
   = \int_{\partial V} dS \left( \check{\sigma}_{ij} u_j \right) - \int_V dV \left( \pdif{\check{\sigma}_{ij}}{x_i} u_j \right) 
\end{equation}
are used. 
$\bol{\check{f}}$ is the dissipative force per unit volume defined as
\begin{equation}
\check{\bol{f}} \equiv \nabla\cdot \check{\bol{\sigma}}^T
\end{equation}
By replacing the time derivative terms with variations, we obtain the non-holonomic constraint as
\begin{equation}
 \int_V dV \left\{ \rho T \delta s - \nu \delta\phi
 -\pdif{x_j}{A_i} \left( \rho T \pdif{s}{x_j} - \nu\pdif{\phi}{x_j} + \check{f}_j \right) \delta A_i \right\} = 0.  \label{App_eq-NH}
\end{equation}

We calculate the stationary condition of the action path \cref{App_eq-action} under the non-holonomic constraint \cref{App_eq-NH}.
The stationary conditions for $K$ and $\beta_i$ give the constraints themselves.
\begin{alignat}{2}
0 &= \frac{\delta I}{\delta  K} = \int_{t_\text{ini}}^{t_\text{fin}} dt \int_V dV \left\{
\rho - \rho_\text{ini} J^{-1} 
\right\} \delta K
& \qquad \to \qquad &
 \rho - \rho_\text{ini} J^{-1} = 0, \\
0 &= \frac{\delta I}{\delta \beta_i} = \int_{t_\text{ini}}^{t_\text{fin}} dt \int_V dV \left\{
\partial_t A_i + \bol{u}\cdot\nabla A_i
\right\} \delta \beta_i
& \qquad \to \qquad &
 \partial_t A_i + \bol{u}\cdot\nabla A_i = 0,
\end{alignat}
Taking the variations for $\rho$ and $\bol{u}$ give
\begin{alignat}{2}
 0 &= \frac{\delta I}{\delta \rho} = \int_{t_\text{ini}}^{t_\text{fin}} dt \int_V dV \left\{
 \pdif{\mathcal{L}}{\rho} + K
 \right\} \delta \rho
& \qquad \to \qquad &
   K = -\frac{1}{2} \bol{u}^2 + \epsilon + \frac{p}{\rho}, \label{App_eq-K}  \\
 0 &= \frac{\delta I}{\delta \bol{u}} = \int_{t_\text{ini}}^{t_\text{fin}} dt \int_V dV \left\{
 \pdif{\mathcal{L}}{\bol{u}}
 + \pdif{}{\bol{u}} \left\{ \beta_i \left( \partial_t A_i +\bol{u}\cdot\nabla A_i \right) \right\}
 \right\} \delta \bol{u}
& \qquad \to \qquad &
 \bol{u} + \frac{\beta_i}{\rho} \nabla A_i = 0.  \label{App_eq-v_b_A}
\end{alignat}
For the variations of $s$, $A_i$ and $\phi$ we obtain
\begin{equation}
\begin{split}
 0 &= \delta I[s,A_i,\phi] \\
   &= \int_{t_\text{ini}}^{t_\text{fin}} dt \int_V dV \left\{
 \pdif{\mathcal{L}}{s}\delta s + \pdif{\mathcal{L}}{\phi}\delta \phi
- \left( \partial_t \beta_i + \nabla\cdot\left(\beta_i\bol{u}\right)  \right)  \delta A_i 
- K J^{-1} \pdif{\rho_\text{ini}}{A_i} \delta A_i
+ \pdif{}{x_j} \left( K \rho_\text{ini} \pdif{J^{-1}}{\left( \partial A_i / \partial x_j \right) } \right) \delta A_i
 \right\} \
 \end{split}
\end{equation}
Using the relations
\begin{equation}
\pdif{J^{-1}}{\left( \partial A_i / \partial x_j \right) } = J^{-1} \pdif{x_j}{A_i}, \qquad
\pdif{}{x_j} \left(\pdif{J^{-1}}{\left( \partial A_i / \partial x_j \right) } \right) = 0,
\end{equation}
the following can be obtained:
\begin{equation}
\begin{split}
& - K J^{-1} \pdif{\rho_\text{ini}}{A_i}
+ \pdif{}{x_j} \left( K \rho_\text{ini} \pdif{J^{-1}}{\left( \partial A_i / \partial x_j \right) } \right) \\
= & - K J^{-1} \pdif{\rho_\text{ini}}{A_i}
+ \pdif{J^{-1}}{\left( \partial A_i / \partial x_j \right) }  \pdif{}{x_j} \left( K \rho_\text{ini} \right) 
+ K \rho_\text{ini}  \pdif{}{x_j} \left(\pdif{J^{-1}}{\left( \partial A_i / \partial x_j \right) } \right) \\
= & - K J^{-1} \pdif{\rho_\text{ini}}{A_i}
+ J^{-1} \pdif{x_j}{A_i} \left( K \pdif{\rho_\text{ini}}{x_j} + \rho_\text{ini}\pdif{K}{x_j}  \right) 
= \rho_\text{ini} J^{-1} \pdif{x_j}{A_i} \pdif{K}{x_j} = \rho \pdif{x_j}{A_i} \pdif{K}{x_j}
 \end{split},
\end{equation}
which gives us
\begin{equation}
\begin{split}
 0 &= \delta I[s,A_i,\phi] \\
   &= \int_{t_\text{ini}}^{t_\text{fin}} dt \int_V dV \left\{
 \pdif{\mathcal{L}}{s}\delta s + \pdif{\mathcal{L}}{\phi}\delta \phi
- \left(
\left( \partial_t \beta_i + \nabla\cdot\left(\beta_i\bol{u}\right)  \right) 
 - \rho \pdif{x_j}{A_i} \pdif{K}{x_j} \right) \delta A_i
 \right\} \\
   &= \int_{t_\text{ini}}^{t_\text{fin}} dt \int_V dV \left\{
 -\rho T\delta s - \rho\mu\delta\phi
- \left(
\left( \partial_t \beta_i + \nabla\cdot\left(\beta_i\bol{u}\right)  \right) 
 - \rho \pdif{x_j}{A_i} \pdif{K}{x_j} \right) \delta A_i
 \right\}.
 \end{split}
\end{equation}
Applying the non-holonomic constraints, we obtain the following.
\begin{gather}
\rho\mu + \nu = 0, \label{App_eq-rhoMuNu}  \\
\pdif{\beta_i}{t} + \nabla\cdot\left( \beta_i \bol{u} \right)
= \pdif{x_j}{A_i} \left( \rho\pdif{K}{x_j} - \rho T \pdif{s}{x_j} + \nu\pdif{\phi}{x_j} - \check{f}_j  \right)
\label{App_eq-beta2}
\end{gather}

By applying an operation $\partial_t + \nabla (\bol{u}\cdot ) -\bol{u}\times\nabla\times$
to \cref{App_eq-v_b_A}, we obtain
\begin{equation}
 \begin{split}
 \partial_t\bol{u} & + \nabla\left(\bol{u}^2\right) - \bol{u}\times\nabla\times\bol{u}
 =  -\left(\partial_t + \nabla (\bol{u}\cdot ) -\bol{u}\times\nabla\times \right)
 \left( \frac{\beta_i}{\rho} \nabla A_i \right) \\
&=  -\partial_t\left( \frac{\beta_i}{\rho} \nabla A_i \right)
 -\nabla\left(\bol{u}\cdot \left(\frac{\beta_i}{\rho} \nabla A_i \right) \right)
 + \bol{u}\times\nabla\times \left(\frac{\beta_i}{\rho} \nabla A_i\right) \\
&=  -\nabla A_i \partial_t\left( \frac{\beta_i}{\rho} \right)
 -\frac{\beta_i}{\rho}  \partial_t\left( \nabla A_i \right)
 -\left(\bol{u}\cdot\nabla A_i \right) \nabla \left(\frac{\beta_i}{\rho} \right)
 -\nabla\left(\bol{u}\cdot \left(\frac{\beta_i}{\rho} \nabla A_i \right) \right)  \\
& \qquad 
 + \bol{u}\times \left( \frac{\beta_i}{\rho} \left( \nabla A_i \right)
 + \nabla\left( \frac{\beta_i}{\rho} \right) \times \nabla A_i \right) \\ 
&=  -\frac{\beta_i}{\rho} \nabla\left\{ \partial_t A_i
 -\left(\bol{u}\cdot\nabla\right) A_i   \right\}
 - \partial_t \left( \frac{\beta_i}{\rho} \right) \nabla A_i 
 - \left(\bol{u}\cdot\left( \nabla A_i \right) \right) \nabla \left( \frac{\beta_i}{\rho} \right)
 + \bol{u}\times \left( \nabla\left( \frac{\beta}{\rho} \right) \times \nabla A_i \right) \\
&=  -\frac{\beta_i}{\rho} \nabla\left\{ D_t A_i  \right\}
 - \partial_t \left(\frac{\beta_i}{\rho} \right) \nabla A_i 
 - \left( \bol{u}\cdot\left(\nabla \left( \frac{\beta_i}{\rho} \right) \right) \right) \nabla A_i \\
&=  -\frac{\beta_i}{\rho} \nabla\left\{ D_t A_i  \right\}
 - D_t \left(\frac{\beta_i}{\rho} \right) \nabla A_i \\
\partial_t\bol{u} & + \frac{1}{2}\nabla\left(\bol{u}^2\right) + \left(\bol{u}\cdot\nabla\right)\bol{u}
=  -\frac{\beta_i}{\rho} \nabla\left\{ D_t A_i  \right\}
 -\left( \frac{1}{\rho} D_t \beta_i - \beta_i \rho^{-2} D_t \rho \right)  \nabla A_i,
\end{split}
\label{App_eq-beta1}
\end{equation}
where the following identities in vector operations are used
\begin{gather}
\frac{1}{2}\nabla\left( \bol{u}^2 \right) - \bol{u}\times\nabla\times\bol{u} = \left(\bol{u}\cdot\nabla\right)\bol{u}, \\
\left( \bol{u}\cdot\nabla a \right) \nabla b
- \bol{u} \times \left( \nabla b \times \nabla a\right)
= \left( \bol{u}\cdot\left(\nabla b\right) \right) \nabla a.
\end{gather}
From $D_t A_i = 0$ (\cref{App_eq-A}) and mass conservation
\begin{equation}
    \partial_t \rho + \left(\bol{u}\cdot\nabla \right) \rho = 0,
\end{equation}
\Cref{App_eq-beta1} can be further simplified as
\begin{equation}
\partial_t\bol{u} + \frac{1}{2}\nabla\left(\bol{u}^2\right) + \left(\bol{u}\cdot\nabla\right)\bol{u}
=  -\frac{1}{\rho} D_t \beta_i  \nabla A_i,
\end{equation}
Using \cref{App_eq-beta2}, we obtain
\begin{equation}
\begin{split}
\partial_t\bol{u} + \frac{1}{2}\nabla\left(\bol{u}^2\right) + \left(\bol{u}\cdot\nabla\right)\bol{u}
&=  -\frac{1}{\rho} 
\pdif{x_j}{A_i} \left( \rho\pdif{K}{x_j} - \rho T \pdif{s}{x_j} + \nu\pdif{\phi}{x_j} - \check{f}_j  \right)
 \pdif{A_i}{x_k} \\
&=  -\frac{1}{\rho} 
\left( \rho\pdif{K}{x_k} - \rho T \pdif{s}{x_k} + \nu\pdif{\phi}{x_k} - \check{f}_k  \right).
\end{split}
\end{equation}
From \cref{App_eq-K}, the following relation is obtained
\begin{equation}
    \pdif{K}{x_j} = -\frac{1}{2} \pdif{\bol{u}^2}{x_j} + T\pdif{s}{x_j} + \frac{1}{\rho}\pdif{p}{x_j},
\end{equation}
which leads
\begin{equation}
\begin{split}
\rho \left(
\partial_t\bol{u} + \frac{1}{2}\nabla\left(\bol{u}^2\right) + \left(\bol{u}\cdot\nabla\right)\bol{u}
\right)
&=  -\rho \left(
-\frac{1}{2} \pdif{\bol{u}^2}{x_j} + T\pdif{s}{x_j} + \frac{1}{\rho}\pdif{p}{x_j}
\right)
+ \rho T \pdif{s}{x_j} - \nu\pdif{\phi}{x_j} + \check{f}_j, \\
\rho \left(
\partial_t\bol{u}  + \left(\bol{u}\cdot\nabla\right)\bol{u}
\right)
&=  - \pdif{p}{x_j} - \nu\pdif{\phi}{x_j} + \check{f}_j, 
\end{split}
\end{equation}
Using $\rho\mu + \nu = 0$ (\cref{App_eq-rhoMuNu}), we obtain the momentum equation as
\begin{equation}
\rho \left( \partial_t\bol{u}  + \left(\bol{u}\cdot\nabla\right)\bol{u} \right)
=  - \nabla p + \rho\mu \nabla\phi + \check{\bol{f}}.   \label{App_eq-NS1}
\end{equation}

\subsection*{Governing equations for model discovery}
In \cref{App_eq-NS1}, $\rho\mu$ can be replaced with the free energy functional,
and the dissipation force $\check{\bol{f}}$ can be written as functional derivative of dissipation potential as
\begin{equation}
\rho \left\{ \partial_t \bol{u} + \bol{u}\cdot\nabla \bol{u} \right\}
= -\nabla p +  \frac{\delta G}{\delta \phi}\nabla\phi  - \frac{\delta \Theta}{\delta \bol{u}},
\end{equation}
where the second term on the right-hand side can be regarded as the reversible force caused by the gradient of the field $\phi$.

Next, we consider the evolution of $\phi$.
Since the dissipation function $\theta$ is now expressed as
\begin{equation}
 \theta(\bol{u}, \phi) 
 = \check{\sigma}_{ij} e_{ij} - \rho\mu D_t \phi
 = \check{\sigma}_{ij} e_{ij} + \mu \nabla\cdot\bol{j}_a,
\end{equation}
which leads to entropy evolution \cref{App_eq-constraintNH} as
\begin{equation}
\begin{split}
 \rho T D_t s &=
 \check{\sigma}_{ij} e_{ij} + \mu \nabla\cdot\bol{j}_a 
 - \nabla\cdot \left( \bol{q} + \mu\bol{j}_a \right) \\
& = \check{\sigma}_{ij} e_{ij} 
 + \mu \nabla\cdot\bol{j}_a 
 - \nabla\cdot \bol{q}
 - \bol{j}_a \cdot \nabla\mu - \mu \nabla\cdot\bol{j}_a \\
&= \check{\sigma}_{ij} e_{ij} 
 - \nabla\cdot \bol{q} - \bol{j}_a \cdot \nabla\mu .
 \end{split}
\end{equation}
From the second law of thermodynamics,  $\bol{j}_a \cdot \nabla\mu \leq 0$ must be satisfied.
Assuming the linear phenomenological law, we can express $\bol{j}_a$ as
\begin{equation}
 \bol{j}_a = -\frac{D}{T} \nabla \mu,
\end{equation}
where $D (>0)$ is a positive constant called a diffusion coefficient.
With the expression of $\mu$ with the free energy functional, the transport equation for $\phi$ can be written as
\begin{equation}
\partial_t \phi + \bol{u}\cdot\nabla\phi
= \nabla\cdot \left( \frac{D}{T} \nabla \frac{\delta G}{\delta \phi} \right).
\end{equation}

The functional derivatives for free energy and dissipation potentials can be generally written as
\begin{equation}
\frac{\delta G}{\delta \phi} = \pdif{g}{\phi}
  - \nabla\cdot\left(\pdif{g}{\left(\nabla\phi \right)} \right)
  + \nabla^2\left(\pdif{g}{\left(\nabla^2\phi \right)} \right) \cdots
  + (-1)^n \nabla^{(n)}\cdot\left(\pdif{g}{\left(\nabla^{(n)}\phi \right)} \right),
\end{equation}
where $n$ can be truncated up to 1 or 2 for the most natural processes.

\addcontentsline{toc}{section}{\protect\numberline{}References}
\bibliographystyle{elsarticle-num-names}
\bibliography{VFD_references}

\end{document}